\documentclass[12pt]{article}

\usepackage{amsfonts,hyperref}
\usepackage{color}
\usepackage{appendix}
\usepackage{amsmath}
\usepackage{graphicx}
\usepackage[latin1]{inputenc}
\usepackage[french,english]{babel}
\addto\captionsfrench{}
\addto\captionsenglish{}

\newcommand\encadremath[1]{\vbox{\hrule\hbox{\vrule\kern8pt
\vbox{\kern8pt \hbox{$\displaystyle #1$}\kern8pt}
\kern8pt\vrule}\hrule}}
\def\enca#1{\vbox{\hrule\hbox{
\vrule\kern8pt\vbox{\kern8pt \hbox{$\displaystyle #1$}
\kern8pt} \kern8pt\vrule}\hrule}}

\newcommand\figureframex[3]{
\begin{figure}[bth]
\hrule\hbox{\vrule\kern8pt
\vbox{\kern8pt \vbox{
\begin{center}
{\mbox{\epsfxsize=#1.truecm\epsfbox{#2}}}
\end{center}
\caption{#3}
}\kern8pt}
\kern8pt\vrule}\hrule
\end{figure}
}
\makeatletter
\@addtoreset{equation}{section}
\makeatother
\newtheorem{theorem}{Theorem}[section]
\newtheorem{conjecture}{Conjecture}[section]
\newtheorem{remark}{Remark}[section]
\newtheorem{proposition}{Proposition}[section]
\newtheorem{lemma}{Lemma}[section]
\newtheorem{corollary}{Corollary}[section]
\newtheorem{definition}{Definition}[section]

\def\br{\begin{remark}\rm\small}
\def\er{\end{remark}}
\def\bt{\begin{theorem}}
\def\et{\end{theorem}}
\def\bd{\begin{definition}}
\def\ed{\end{definition}}
\def\bp{\begin{proposition}}
\def\ep{\end{proposition}}
\def\bl{\begin{lemma}}
\def\el{\end{lemma}}
\def\bc{\begin{corollary}}
\def\ec{\end{corollary}}
\def\beaq{\begin{eqnarray}}
\def\eeaq{\end{eqnarray}}

\newcommand{\beq}{\begin{equation}}
\newcommand{\eeq}{\end{equation}}
\newcommand{\bea}{\begin{eqnarray}}
\newcommand{\eea}{\end{eqnarray}}

 \newcommand{\Tr}{{\,\rm Tr}\:}

\newcommand{\td}[1]{{\tilde{#1}}}

\newcommand{\om}{\omega}

\newcommand{\ii}{{\mathrm{i}}}
\newcommand{\e}{{\,\rm e}\,}
\newcommand{\ee}[1]{{{\rm e}^{#1}}}

\newcommand{\Pint}{{\int\kern -1.em -\kern-.25em}}

\newcommand{\genus}{{\mathfrak{g}}}

\newcommand\Res{\mathop{{\rm Res}}}

\begin{document}
%=============================Page de titre==============%\date{??}

\sloppy

%\maketitle

\pagestyle{empty}
\hfill SPhT-T14/033, CRM-3335
\addtolength{\baselineskip}{0.20\baselineskip}
\begin{center}
\vspace{26pt}
{\large \bf {
A short overview of the "Topological recursion"\\
%A From random matrices to geometry\\
%How emerged the "topological recursion"\\
%ICM 2014
}}
\vspace{20pt}
%\newline
\textbf{B. Eynard}\\
Institut de Physique Théorique,
CEA, IPhT, F-91191 Gif-sur-Yvette, France
CNRS, URA 2306, F-91191 Gif-sur-Yvette, France
,\\
CRM Centre de Recherches Math\'ematiques de Montr\'eal.
\end{center}

\vspace{20pt}
\begin{center}
{\bf Abstract}

This is the long version of the ICM2014 proceedings.
It consists in a short overview of the "topological recursion", a relation appearing in the asymptotic expansion of many integrable systems and in enumerative problems.
We  recall how computing large size asymptotics in random matrices, has allowed to discover some fascinating and ubiquitous geometric invariants.
Specializations of this method recover many classical invariants, like Gromov--Witten invariants, or knot polynomials (Jones, HOMFLY,...).
In this short review, we give some examples, give definitions, and review some properties and applications of the formalism.
%This method is closely related to mirror symmetry.

\end{center}

%\tableofcontents

%-----------------------------ABSTRACT--------------------------------------

\vspace{26pt}
\pagestyle{plain}
\setcounter{page}{1}

%*********************************************************************
%==================== ARTICLE =======================================
%******************************************

\section{Introduction}

The "topological recursion" is a recursive definition (axiomatic definition in \cite{EOFg, EOreview}), which associates a double family (indexed by two non--negative integers $g$ and $n$) of differential forms $\omega_{g,n}$, to a "spectral curve" ${\cal S}$ (a plane analytical curve with some additional structure, see definition below).

The $\omega_{g,n}({\cal S})$'s are  called the "invariants" of the spectral curve ${\cal S}$.

$$
{\rm Topological\,\, Recursion:}\qquad
{\rm spectral\,curve}\,\,{\cal S}
\,\,\longrightarrow\,\,
{\rm Invariants}\,\,\, \omega_{g,n}({\cal S})
$$

\medskip

The initial terms $\omega_{0,1}$ and $\omega_{0,2}$ are some canonical 1-form and 2-form on the spectral curve ${\cal S}$, the other $\omega_{g,n}$'s are then defined by a universal recursion on $(2g+n-2)$.
$\omega_{g,n}({\cal S})$ is a symmetric $n-$form on ${\cal S}^n$, and the $n=0$ invariant, customarily denoted $F_g({\cal S}) =\omega_{g,0}({\cal S})$, is a number $F_g({\cal S})\in\mathbb C$ (or in fact an element of the field over which ${\cal S}$ is defined).

\medskip

Those invariants have fascinating mathematical properties, they are "symplectic invariants" (invariants under some symplectic transformations of the spectral curve), they are almost modular forms (under the modular $Sp_{2\genus}(\mathbb Z)$ group when the spectral curve has genus $\genus$), they satisfy Hirota-like equations, they satisfy some form-cycle duality deformation relations (generalization of Seiberg-Witten relations), they are stable under many singular limits, and enjoy many other fascinating properties...

\medskip

Moreover, specializations of those invariants recover many known invariants, including volumes of moduli spaces, Hurwitz numbers, intersection numbers, Gromov-Witten invariants, numbers of maps (Tutte's enumeration of maps), or asymptotics of random matrices expectation values.
And since very recently, it is conjectured that they also include knot polynomials (Jones, HOMFLY, super polynomials...), which provides an extension of the volume conjecture.

\medskip

The purpose of this short article is only a small glimpse of the fast evolving mathematics of those invariants.

We shall present here a few examples, then mention how these invariants were first discovered in random matrix theory, and then observed or conjectured in many other areas of maths and physics.

\section{Examples of topological recursions}

In this section, we jsut show some examples where a "topological recursion" structure was known.

The purpose of the present section is to give some examples, coming from apparently unrelated geometric problems, and which show a similar recursive structure.
We shall thus remain very introductory, and don't give precise definitions (these can be found in the literature).

Since the full general definition of the "topological recursion and symplectic invariants" requires a substantial Riemannian geometry background, not needed in those introductory examples, we postpone the precise definition of the topological recursion to section \ref{secdeftoprec}, and we first focus on examples.

\subsection{Mirzakhani's recursion for hyperbolic volumes}

\begin{figure}
\noindent\framebox{
\vbox
{$$ \includegraphics[scale=0.3]{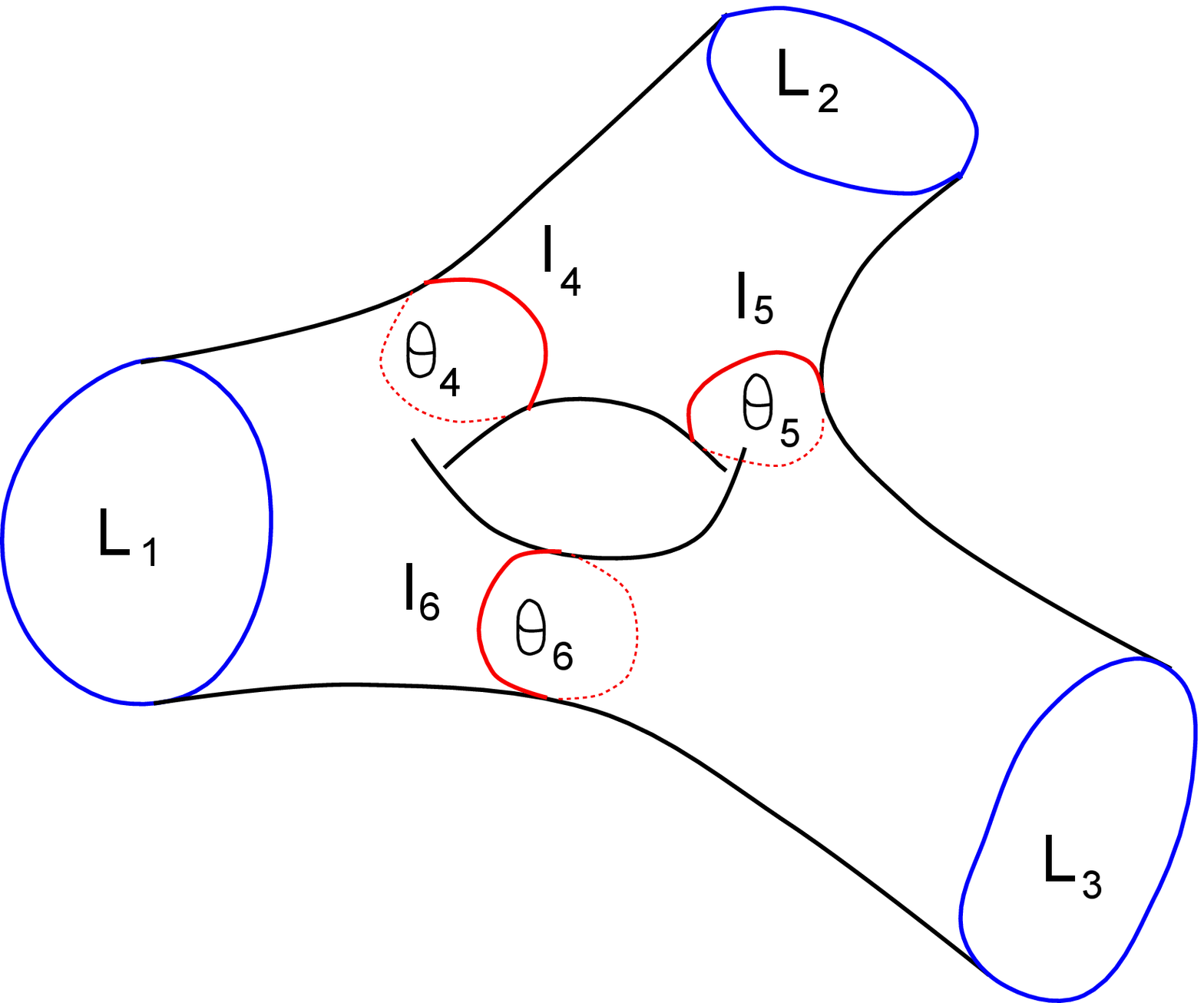}$$
\em The Weil-Petersson metrics and Fenchel-Nielsen coordinates on ${\cal M}_{g,n}$ are obtained as follows: let $2g-2+n>0$, let $L_1,\dots,L_n$ be positive real numbers, and let $\Sigma\in{\cal M}_{g,n}$. The Poincaré metrics on $\Sigma$ is the unique metrics of constant negative curvature $-1$, such that the boundaries of $\Sigma$ are geodesics of repsective lengths $L_1,\dots,L_N$.
Then, we may cut $\Sigma$ into $2g-2+n$ pairs of pants, all of whose boundaries are geodesics. This cutting is not unique. Vice versa, a connected gluing of $2g-2+n$ hyperbolic pairs of pants along their geodesic boundaries, gives a unique Riemann surface in ${\cal M}_{g,n}$. Boundaries of pairs of pants can be glued together provided that the glued geodesics have the same lengths, and they can be rotated by some angle.
The $3g-3+n$ lengths of the glued boundaries and the $3g-3+n$ gluing angles, are called the Fenchel Nielsen coordinates. They are local coordinates on ${\cal M}_{g,n}$. They are not defined globally because of non-unicity of the cutting.
However, the form $w=\prod_i d\ell_i\wedge d\theta_i$, called the Weil-Petersson form, is globally defined. The Weil--Petersson volume is
${\cal V}_{g,n}(L_1,\dots,L_n) = \int_{{\cal M}_{g,n}} w$.}}
\caption{Weil-Petersson volumes in a nut shell\label{figWPdef}}
\end{figure}

See short definition in fig.\ref{figWPdef}.

Let $g,n$ be non--negative integers such that $2g-2+n>0$ (i.e. $(g,n)=(0,0),(0,1),(0,2),(1,0)$ are excluded).
Let ${\cal V}_{g,n}(L_1,\dots,L_n)$ be the hyperbolic volume (called "Weil-Petersson volume" \cite{Mirza1} ,see fig \ref{figWPdef}) of the moduli-space ${\cal M}_{g,n}$ of  genus $g$ bordered Riemann surfaces with $n$ geodesic boundaries of respective lengths $L_1,\dots,L_n$
$$
{\cal V}_{g,n}(L_1,\dots,L_n) = \int_{{\cal M}_{g,n}\,, \, \, \ell(\partial_i)=L_i}\,w\, ,
\qquad\quad  \,{\rm where}\,\,\,
w={\rm Weil-Petersson\,form},
$$
and let its Laplace transform:
$$
W_{g,n}(z_1,\dots,z_n) = \int_0^\infty\dots \int_0^\infty {\cal V}_{g,n}(L_1,\dots,L_n)\,\,\prod_{i=1}^n \e^{-z_i L_i}\,L_i dL_i
$$
Those hyperbolic volumes, are not easy to compute with hyperbolic geometry.
Only the first of them (smallest values of $g$ and $n$) had been computed directly by hyperbolic geometry, for example:
$$
{\cal V}_{0,3}(L_1,L_2,L_3)=1 \qquad , \quad W_{0,3}(z_1,z_2,z_3)=\frac{1}{z_1^2\,z_2^2\,z_3^2}
$$
$$
{\cal V}_{1,1}(L)=\frac{1}{48}\,(L^2+4\pi^2) \qquad , \quad W_{1,1}(z)=\frac{1}{8 z^4} + \frac{\pi^2}{12\,z^2}
$$
It turns out (this is not obvious from the definition) that these volumes are even polynomials of the $L_i$'s, or equivalently, the $W_{g,n}$'s are even polynomials of the $1/z_i$'s.

In 2004, M. Mirzakhani discovered a beautiful recursion relation \cite{Mirza1}, which computes all volumes ${\cal V}_{g,n}$ for all $g$ and $n$, by recursion on $2g+n$. We shall not write Mirzakhani's relation among the ${\cal V}_{g,n}$'s, but we shall consider here its Laplace transformed version:
\bt[Topological recursion for Weil-Petersson volumes] {\bf Mirzakhani's recursion \cite{Mirza1}, Laplace transformed\cite{EOWP}}

For any $(g,n)$ such that $2g-2+n>0$, one has:
\bea
W_{g,n}(z_1,\overbrace{z_2,\dots,z_n}^{J})
&=& \Res_{z\to 0} \frac{1}{(z_1^2-z^2)}\,\frac{\pi}{\sin{(2\pi z)}}\,\,\,\Big[ W_{g-1,n+1}(z,-z,J) \cr
&& +\sum_{I\uplus I'=J; h+h'=g}' W_{h,1+\# I}(z,I)\,W_{h',1+\# I'}(-z,I') \Big]\,dz \cr
\eea
where $\sum'$ means that we exclude from the sum the two cases $(I=J,h=g)$ and $(I'=J,h'=g)$, and we have denoted:
\beq
W_{0,2}(z_1,z_2) = \frac{1}{(z_1-z_2)^2}.
\eeq

\et

\smallskip

This theorem \cite{mulsaf, Xu, Mulase2006,EOWP} is very efficient at actually computing the volumes.
It is a recursion on the Euler--characteristics $\chi=2-2g-n$, at each step, the absolute value of the Euler characteristics in the Left Hand Side, is one more than the total Euler characteristics of every Right Hand Side terms:
$$
|2-2(g-1)-(n+1)| = |2-2h-(1+m)+2-2(g-h)-(1+n-1-m)| = |2-2g-n|-1.
$$
This explains the name "{\bf topological recursion}".

Let us illustrate the theorem for the case $(g,n)=(1,1)$. For $(g,n)=(1,1)$ it says that:
$$
W_{1,1}(z_1)
= \Res_{z\to 0} \frac{1}{(z_1^2-z^2)}\,\frac{\pi}{\sin{(2\pi z)}}\,\,\, W_{0,2}(z,-z)\,\,dz
$$
Then let us compute the residue:
\bea
W_{1,1}(z_1)
&=& \Res_{z\to 0} \frac{1}{(z_1^2-z^2)}\,\frac{\pi}{\sin{(2\pi z)}}\,\,\, W_{0,2}(z,-z)\,\,dz \cr
&=& \Res_{z\to 0} \frac{1}{z_1^2\,(1-\frac{z^2}{z_1^2})}\,\frac{\pi}{2\pi \,z(1-\frac{4\pi^2\,z^2}{6}+O(z^4))}\,\,\, \frac{1}{4\, z^2}\,\,dz \cr
&=& \frac{1}{8\,z_1^2}\,\Res_{z\to 0} \frac{dz}{z^3}\,\,(1-\frac{z^2}{z_1^2}+O(z^4))^{-1}\,\,(1-\frac{4\pi^2\,z^2}{6}+O(z^4))^{-1} \cr
&=& \frac{1}{8\,z_1^2}\,\Res_{z\to 0} \frac{dz}{z^3}\,\,(1+\frac{z^2}{z_1^2}+\frac{4\pi^2\,z^2}{6}+O(z^4) ) \cr
&=& \frac{1}{8\,z_1^2}\,\left(\frac{1}{z_1^2}+\frac{4\pi^2}{6}\right) \cr
\eea
which coincides with the result previously known from hyperbolic geometry.

\medskip

This theorem is an illustration of a universal and far more general recursive structure, called the "topological recursion", as we shall see below.

\subsection{Hurwitz numbers}

\begin{figure}
\noindent\framebox{
\vbox
{\em
$$\includegraphics[scale=0.45]{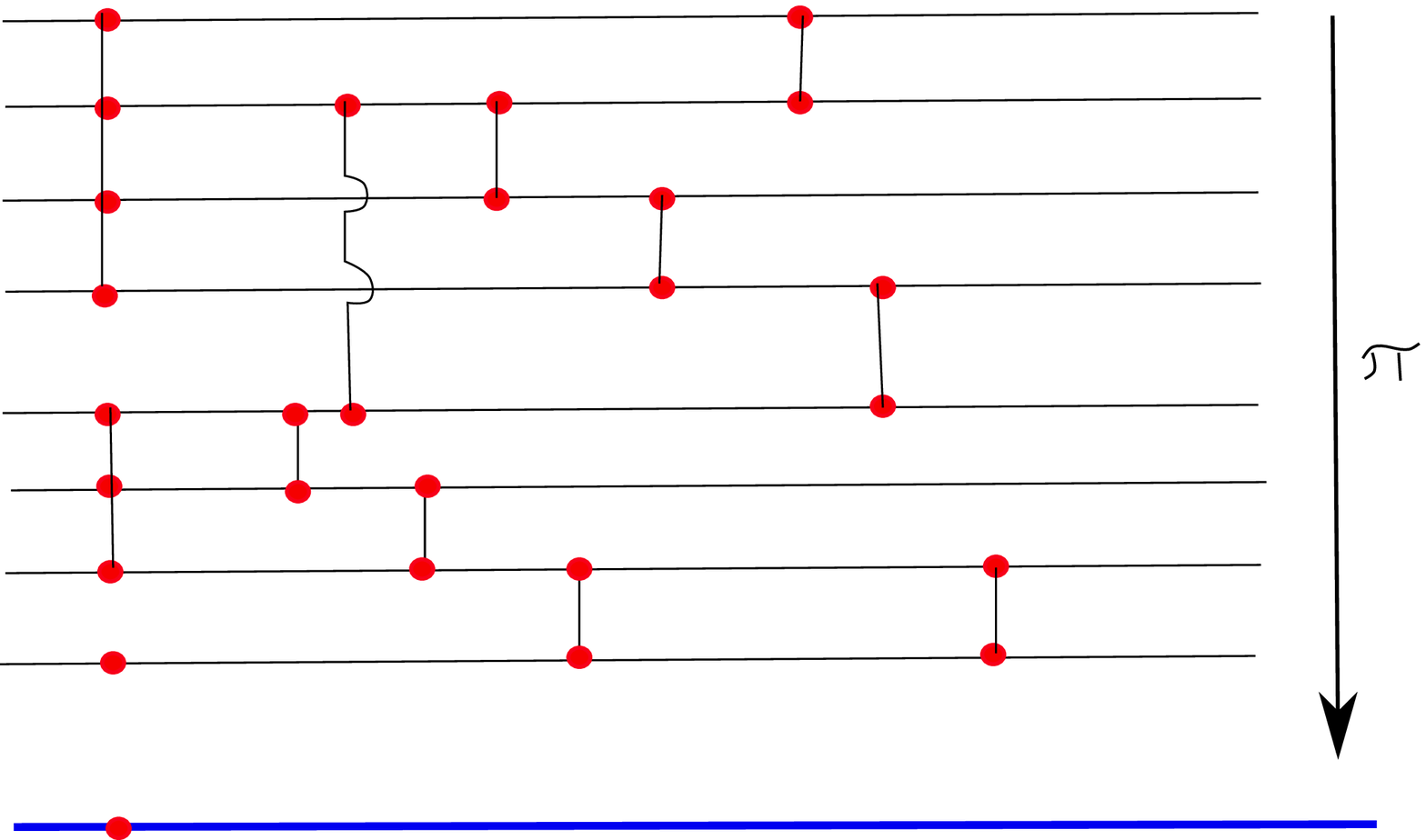}$$
A ramified covering $(\Sigma,\pi)$ of the Riemann sphere $\mathbb CP^1$, is the data of a Riemann surface $\Sigma$, and an analytical map $\pi:\Sigma\to\mathbb CP^1$ of some degree $d$.
For all generic points $x\in \mathbb CP^1$, the preimage $\pi^{-1}(x)\in\Sigma$ consists of $d$ points $\# \pi^{-1}(x)=d$.
Branchpoints are  the points $x\in\mathbb CP^1$ such that $\# \pi^{-1}(x)<d$. Ramification points are the preimages of branchpoints.

%Near a generic point $p\in \Sigma$, the map $\pi:\Sigma\to \mathbb CP^1$ is analytical and locally invertible.

Ramification points are the points near which the map $\pi:\Sigma\to \mathbb CP^1$ is analytical but not locally invertible.
A ramification point $a$ is said of degree $r=\deg(a)$, if locally near $a$, the map $\pi$ behaves like (in any choice of local coordinate)
$$
\pi: p\mapsto \pi(a) + c_a (p-a)^r + O((p-a)^{r+1})\qquad , \,\,\, c_a\neq 0
$$

Let $x$ be a branchpoint, and $\{a_1,\dots,a_l\}=\pi^{-1}(x)$ be its preimages on $\Sigma$, and let $r_i=\deg a_i$ be its degrees. We assume that we have ordered the points $a_i$'s such that $r_1\geq r_2\geq \dots \geq r_l$.
Then $(r_1,\dots,r_l)$ is called the ramification profile of $x$.

A regular branchpoint $x$ is of degree $2$, and such that $\#\pi^{-1}(x)=d-1$, its ramification profile is  $(2,\overbrace{1,\dots,1}^{d-2})$.

The Simple Hurwitz numbers $H_g(\mu)$ count the number of (equivalence homotopy classes) ramified coverings $(\Sigma,\pi)$ such that $\Sigma$ is a connected surface of genus $g$, and $\pi$ has only one non--regular branchpoint, whose profile is given by the partition $\mu=(\mu_1,\dots,\mu_l)$.

The Hurwitz formula implies that such a covering must have $b=2g-2+\sum_i (\mu_i+1))$ regular branchpoints.
}}
\caption{Hurwitz numbers in a nutshell.\label{figHurwitzdef}}
\end{figure}

See short definition in fig \ref{figHurwitzdef}.

Let $H_{g,n}(\mu)$ be the simple Hurwitz number of genus $g$ with ramification profile $\mu$. In other words, let $H_{g,n}(\mu)$ be the number of connected ramified coverings of the Riemann sphere, of genus $g$ and with only one multiply ramified point whose ramification profile is a partition $\mu=(\mu_1\geq \mu_2\geq\dots\geq \mu_n)$ of length $n$  (we denote $n=\ell(\mu)$ the length of $\mu$, and $|\mu|=\sum_i \mu_i$ its weight), and all other ramification points are simple (
and Riemann  Hurwitz formula says that there are $b=2g-2+n+|\mu|$ simple ramification points).

In other words, $H_{g,n}(\mu)$ is the number of ways to glue $|\mu|$ sheets ($|\mu|$ copies of the Riemann sphere), along cuts open between ramification points, forming a connected surface of genus $g$. Topologically, a ramified covering is the data of the deck transformations (a permutation of sheets) at each branchpoints, requiring that one of the permutations belongs to the conjugacy class given by $\mu$ and all other permutations be transpositions (simple branchpoints), and the product of all of them has to be identity (condition for the gluing to make a smooth surface).

For example when $\ell(\mu)=1$ and genus $g=0$, one has that $H_{0,1}(\mu_1)$ is the number of ways of gluing $\mu_1$ sheets together, at one fully ramified point (partition $\mu=(\mu_1)$), and at $\mu_1-1$ simple ramification points, which make a surface of genus 0, i.e. planar, i.e. without loops. Such a ramified covering is thus the data of $\mu_1$ sheets (represented by $\mu_1$ points) linked by $\mu_1-1$ ramification points (represented by $\mu_1-1$ edges), connected and without loops.
Therefore this is the same thing as counting the number of covering trees which can be drawn on the complete graph with $\mu_1$ points.
This is given by Cayley's \cite{Cayley} formula:
\beq
H_{0,1}(\mu_1) = \pm \det_{\mu_1-1}
\begin{pmatrix}
1-\mu_1 & 1   & 1 &  \dots & 1 \cr
1 & 1-\mu_1  & 1   &  \dots & 1 \cr
1 &  \ddots & \ddots  &   \ddots  &  1 \cr
1 &  & 1 & 1-\mu_1      &  1 \cr
1 & \dots  & 1 & 1  & 1-\mu_1   \cr
\end{pmatrix} = \mu_1^{\mu_1-2}
\eeq
With genus 0 and partitions of length $2$, one finds (though not easily \cite{Goulden2000, BM}):
\beq
H_{0,2}(\mu_1,\mu_2) = (\mu_1+\mu_2-1)!\,\frac{\mu_1^{\mu_1+1}\,\mu_2^{\mu_2+1}}{\mu_1!\,\mu_2!} .
\eeq

Out of Hurwitz numbers, we define the generating functions, which are some kinds of discrete Laplace transforms of the $H_{g,n}$'s:
\beq
W_{g,n}(x_1,\dots,x_n) = \sum_{\mu,\,\ell(\mu)=n} \frac{H_{g,n}(\mu)}{(2g-2+n+|\mu|)!}\,\,\sum_{\sigma\in\mathfrak{S}_n} \prod_{i=1}^n \e^{\mu_i x_{\sigma(i)}}.
\eeq
%where $z_\mu=n!/\prod_i n_i!$ with $n_i=\#\{j,\, \mu_j=i\}$.

For example:
\beq
W_{0,1}(x) = \sum_{k=1}^\infty \frac{k^{k-2}}{(k-1)!}\,\e^{kx}  = L(\e^{x})
\eeq
where $L$ is the Lambert function, i.e. solution of $\e^{x}=L\,\e^{-L}$.
Similarly
\bea
W_{0,2}(x_1,x_2)  &=& \sum_{k,k'=1}^\infty \frac{k^{k+1}\,k'^{k'+1}}{(k+k')\,k!\,k'!}\,\,\e^{kx_1}\,\,\,\e^{k'x_2}  \cr
& =& \frac{L(\e^{x_1})}{1-L(\e^{x_1})}\,\frac{L(\e^{x_2})}{1-L(\e^{x_2})}\,\frac{1}{(L(\e^{x_1})-L(\e^{x_2}))^2} - \frac{\e^{x_1}\,\e^{x_2}}{(\e^{x_1}-\e^{x_2})^2}.
\eea
One finds that it is easier to make a change of variable and work with $z_i=L(\e^{x_i})$ rather than $x_i$, and thus define the following differential forms:
\beq
\omega_{g,n}(z_1,\dots,z_n) = \left(W_{g,n}(x_1,\dots,x_n)\,+\delta_{g,0}\,\delta_{n,2}\,\frac{\e^{x_1}\,\e^{x_2}}{(\e^{x_1}-\e^{x_2})^2}\right)\,dx_1\dots dx_n.
\eeq
For example:
\beq
\omega_{0,1}(z) = (1-z)\,dz
\qquad , \qquad
\omega_{0,2}(z_1,z_2) = \frac{dz_1\,dz_2}{(z_1-z_2)^2}.
\eeq

Goulden, Jackson, Vainshtein \cite{Goulden2000} derived a recursion formula (called "cut and join equation") satisfied by those Hurwitz numbers, and after Laplace transform \cite{EMS}, one finds (not so easily) the topological recursion formula, which was first conjectured by physicists Bouchard and Mari\~no \cite{BM}:
\bt[Topological recursion for Hurwitz numbers]
{\bf = Bouchard-Mari\~no conjecture} (first proof in \cite{EMS}).

The forms $\omega_{g,n}$'s satisfy the following recursion:
\bea
\omega_{g,n}(z_1,z_2,\dots,z_n)
&=& \Res_{z\to 1} K(z_1,z)\,\,\,\Big[ \omega_{g-1,n+1}(z,s(z),z_2,\dots,z_n) \cr
&& +\sum_{I\uplus I'=\{z_2,\dots,z_n\}; h+h'=g}' \omega_{h,1+\# I}(z,I)\,\omega_{h',1+\# I'}(s(z),I') \Big] \cr
\eea
where $\sum'$ means that we exclude from the sum the two cases $(I=\{z_2,\dots,z_n\},h=g)$ and $(I=\emptyset,h=0)$, and where the recursion kernel $K$ is:
\beq
K(z_1,z) = \frac{\,dz_1}{2}\,\frac{\frac{1}{z_1-z}-\frac{1}{z_1-s(z)}}{(z-s(z))}\,\frac{z}{(1-z)\,dz}
\eeq
and where the map $s:z\mapsto s(z)$ defined in a vicinity of $z=1$, is the involution such that  $s\neq {\rm Id}$ and  solution  of
\beq
s(z)\,\e^{-s(z)} = z\,\e^{-z} \quad , \, s(1)=1.
\eeq
Locally near $z=1$, its Taylor series expansion starts with:
\beq
s(z) = 1 - (z-1) + \frac{2}{3} (z-1)^2 - \frac{4}{9} (z-1)^3 + \frac{44}{135} (z-1)^4- \frac{104}{405} (z-1)^5+ \dots
\eeq
\et

This recursion is very efficient at computing, for instance it easily gives:
\beq
\omega_{0,3}(z_1,z_2,z_3) = \frac{dz_1\,dz_2\,dz_3}{(1-z_1)^2\,(1-z_2)^2\,(1-z_3)^2}
\eeq
\beq
\omega_{1,1}(z) = \frac{1}{24}\,\left( \frac{1+2z}{(1-z)^4} - \frac{1}{(1-z)^2}\right)\,dz
\eeq

%In general, for $2g-2+n>0$ the recursion implies that $\omega_{g,n}$ is a polynomial of $t_i=1/(1-z_i)$:

Again, this theorem is an illustration of the universal "{\bf topological recursion}", as we shall see below.

\subsection{Counting maps}

We give another example of a topological recursion in combinatorics.

Counting maps (discrete surfaces), has been a fascinating question since the works of Tutte \cite{tutte, tutte2} in the 60's, when he was able to give explicit formulae for counting planar triangulations or planar quadrangulations.

\begin{figure}
\noindent\framebox{
\vbox{
$$\includegraphics[scale=0.4]{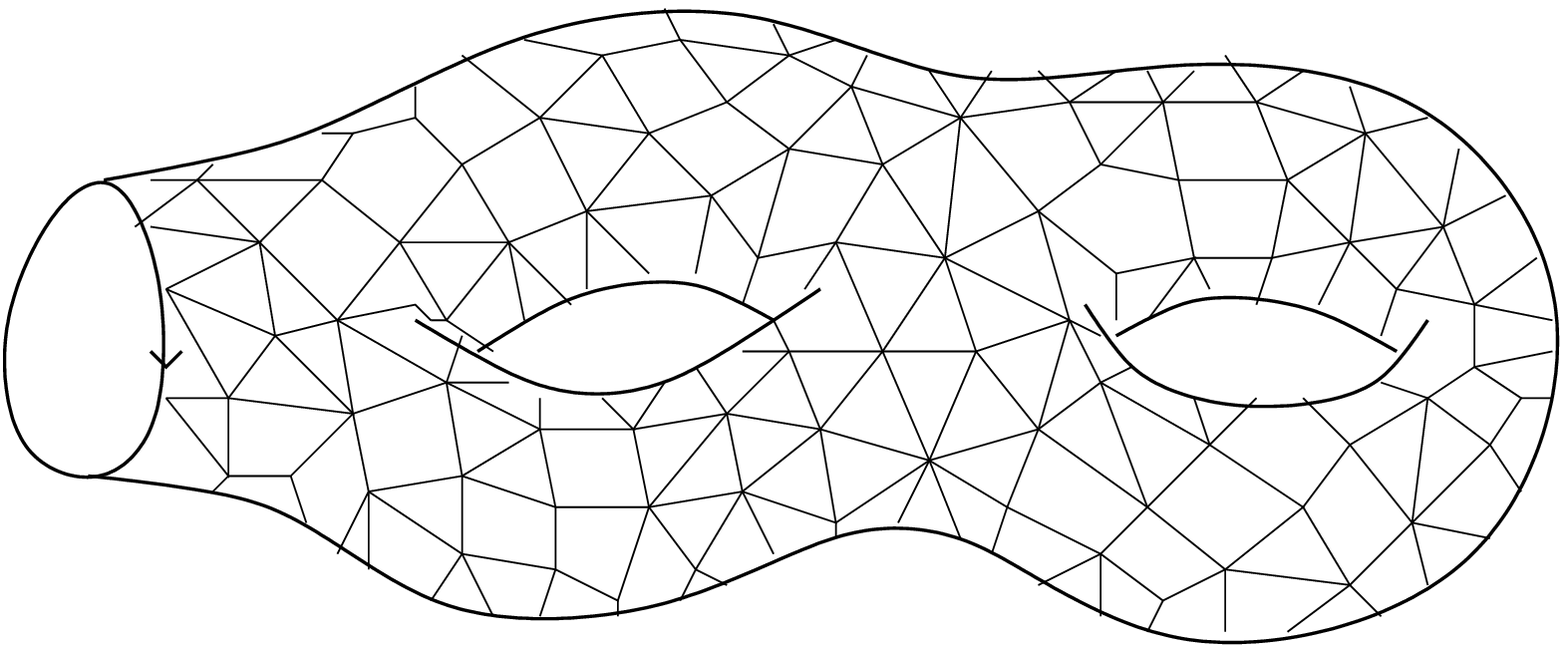}$$
{\em
A map of genus $g$ is a cellular connected graph $G$ embedded on a connected orientable surface $\Sigma$ of genus $g$ (cellular means that $\Sigma\setminus G$ is a disjoint union of topological discs), and modulo all reparametrizations.
The connected components of $\Sigma\setminus G$ are called the faces, they are bordered by edges and vertices of the graph. We say that the size of a face, is the number of edges bordering it.
We shall assume that some faces can be marked faces, and some are unmarked.
We shall always require that unmarked faces have size $\geq 3$, and that marked faces have a marked edge on their boundary (a root).

Let $\mathbb M_{g,n}(v)$ be the set of maps of genus $g$, with $v$ vertices, and $n$ marked faces, and an arbitrary number of unmarked faces (we recall that all unmarked faces have a size $\geq 3$).

The set $\mathbb M_{g,n}(v)$ is a finite set (easy to prove by writing the Euler characteristics).

Here is the example of the first few $\mathbb M_{0,1}(v)$ for $v=1,2,3$ (the marked face is the exterior face of the planar graph)
$$\includegraphics[scale=0.3]{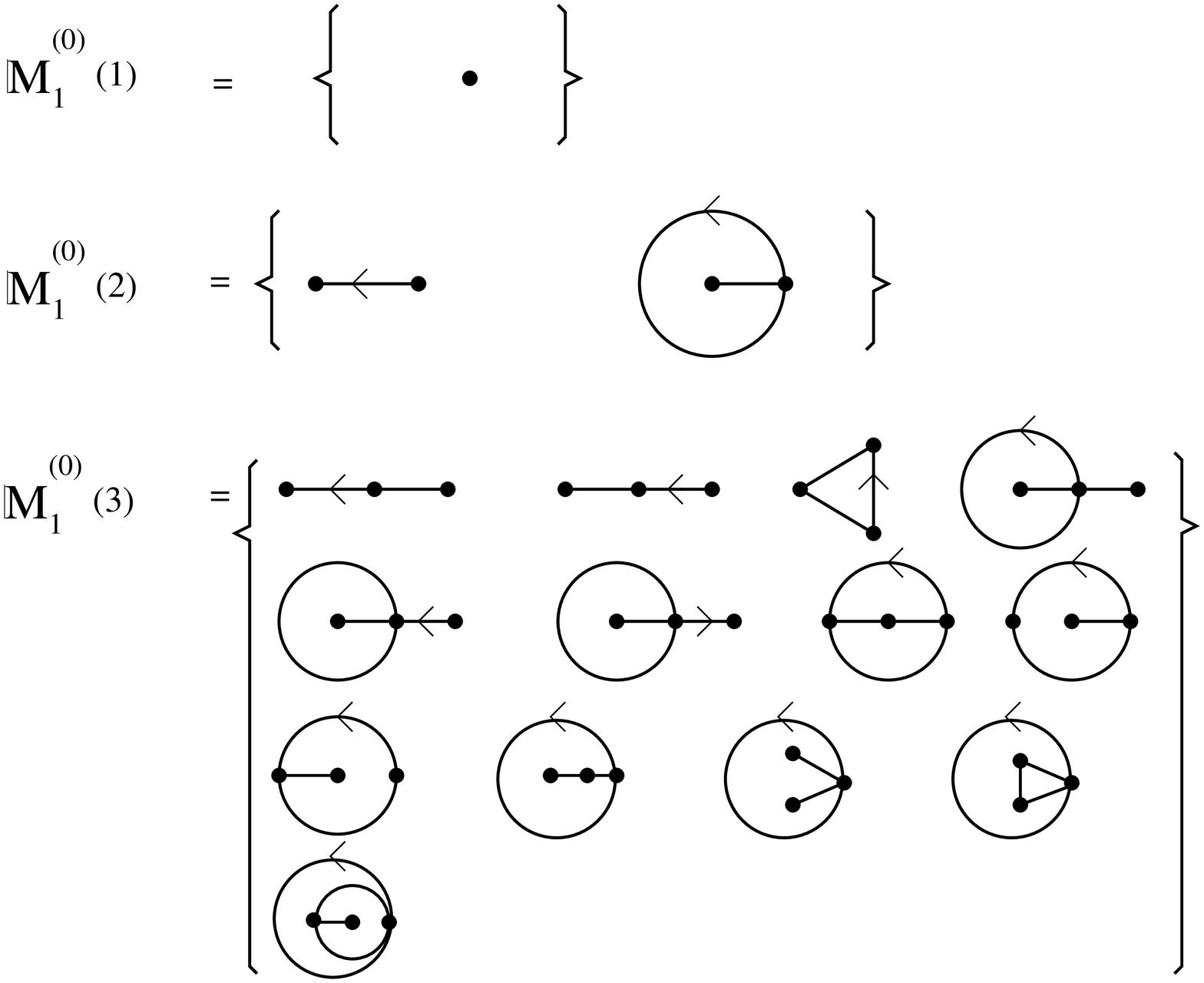}$$

}}}
\caption{Maps in a nutshell.\label{figMapsdef}}
\end{figure}

We shall refer the reader to the literature on maps, or see fig.\ref{figMapsdef}.

\bd
We define the following generating series for counting maps:
\beq
W_{g,n}(x_1,\dots,x_n;t_3,t_4,\dots;t) \in \mathbb Q[1/x_i,t_j][[t]]
\eeq
(i.e. formal series of $t$, whose coefficients are polynomials over $\mathbb Q$ of $1/x_1,1/x_2,\dots,1/x_n, t_3,t_4,\dots$), by:
\beq
W_{g,n} =\frac{t}{x_1}\,\delta_{g,0}\delta_{n,1}+ \sum_{v\geq 1} t^v\, \sum_{m\in \mathbb M_{g,n}(v)} \frac{1}{\#{\rm Aut}(m)}\,\frac{\prod_{j\geq 3} t_j^{n_j(m)}}{\prod_{i =1}^n x_i^{1+l_i(m)}}
\eeq
where $\mathbb M_{g,n}(v)$ is the (finite) set of maps of genus $g$ with $n$ labeled marked faces and $v$ vertices. If $m\in \mathbb M_{g,n}(v)$, $l_i(m)$ is the size of the $i^{\rm th}$ marked face, and $n_j(m)$ is the number of unmarked $j-$gons of $m$, and $\#{\rm Aut}(m)$ is the Automorphism group of the map $m$ (it is always 1 if $n\geq 1$ i.e. if the map is rooted).

\ed

Very often we shall omit to write the dependance in $t$ and the $t_j$'s and keep only the $x_i$'s dependence, because the recursion equations will act on those variables only, and thus write as a shorter notation:
\beq
W_{g,n}(x_1,\dots,x_n) \quad \mathop{\equiv}^{{\rm notation}}\quad W_{g,n}(x_1,\dots,x_n;t_3,t_4,\dots;t) .
\eeq

\medskip

Tutte in the 60's found a recursion formula \cite{tutte, tutte2} for counting maps, by recursively removing the marked edge of the 1st face.
In fact Tutte worked only in $\mathbb M_{0,1}(v)$, i.e. planar maps with one marked edge (rooted planar maps), but his recursion equation straightforwardly extends to arbitrary genus and arbitrary number of marked faces.
Therefore Tutte computed $W_{0,1}$.

For example, for quadrangulations (i.e. set $t_3=t_5=t_6=\dots = 0$ and only $t_4\neq 0$), he found:
\beq\label{TuttequadW01}
W_{0,1}(x) = \frac{1}{2}\,\left( x-t_4 x^3 + t_4 (x^2+2\gamma^2-\frac{1}{t_4})\sqrt{x^2-4\gamma^2} \right)
\,\,\, , \,\,\gamma^2 = \frac{1}{6t_4}\,\left(1-\sqrt{1-12 t t_4}\right).
\eeq
This is equivalent to say that the number of rooted planar quadrangulations with $n$ unmarked faces and one marked face of size $2l$ is:
\beq
3^n\,\, \frac{(2l)!}{l!\,(l-1)!}\,\,\frac{(2n+l-1)!}{(l+n+1)!\, n!}.
\eeq
In particular if the marked face has size $4$ (i.e. $l=2$), we recover the famous Tutte's formula \cite{tutte2} for the number of planar rooted quadrangulations with $m=n+1$ faces (including the marked one):
\beq
\#\,\,{\rm genus}\,0\, , \,m\,{\rm faces}:\qquad \quad
2\,.\,3^{m}\,\, \,\,\frac{(2m)!}{(m+2)!\, m!}.
\eeq
The function $W_{0,1}(x)$ looks better if written parametrically:
\beq
\left\{
\begin{array}{l}
x = \alpha+\gamma(z+1/z) \cr
W_{0,1} = \sum_{k\geq 1} v_k z^{-k} =y(z)\cr
\end{array}
\right.
\eeq
where the parameters $\{\alpha,\gamma, v_1,v_2,v_3,\dots\}$ are a reparametrization of the weights $\{t_3,t_4,\dots, t_k\,\dots\}$ assigned to triangles, quadrangles, $\dots$, k-gons, $\dots$.
More explicitly, the $v_k$'s are functions of $\alpha$ and $\gamma$ and the $t_k$'s by writing
\beq
\sum_k v_k (z^k+z^{-k}) = x-\sum_{k\geq 2} t_{k+1} x^k
\quad , \quad x=\alpha+\gamma(z+1/z),
\eeq
and $\alpha$ and $\gamma$ are determined by $v_0=0$ and $v_1=t/\gamma$.

For example for quadrangulations (i.e. set $t_3=t_5=t_6=\dots = 0$ and only $t_4\neq 0$), only $v_1,v_3,\gamma$ are non--vanishing, and have to satisfy:
$$
v_1(z+z^{-1})+v_3(z^3+z^{-3}) = \gamma(z+z^{-1}) - t_4 \gamma^3 (z+z^{-1})^3
$$
i.e.
$$
v_1=\gamma-3t_4\gamma^3
\quad , \quad
v_3=-t_4\gamma^3,
$$
and the equation $v_1=t/\gamma$ determines $\gamma$:
$$
\frac{t}{\gamma}=\gamma-3t_4\gamma^3
\qquad \Rightarrow \quad
\gamma^2= \frac{1}{6t_4}\,\left(1-\sqrt{1-12 t t_4}\right)
$$
We indeed  recover Tutte's result \eqref{TuttequadW01}, written in variable $z$ (related to $x$ by $x=\gamma(z+1/z)$)
\beq
W_{0,1}(x)=y(z) = \frac{v_1}{z}+\frac{v_3}{z^3} = \frac{t}{\gamma z} - \frac{3 t_4\gamma^3}{z^3} .
\eeq

With  this reparametrization in terms of $z$, we rewrite the generating functions of maps $W_{g,n}(x_1,\dots,x_n)$, in terms of the variables $z_i$, and as differential forms:
\beq
\omega_{g,n}(z_1,\dots,z_n) = W_{g,n}(x_1,\dots,x_n)\,dx_1\dots dx_n + \delta_{g,0}\delta_{n,2}\,\frac{dx_1\,dx_2}{(x_1-x_2)^2} \qquad , x_i=x(z_i).
\eeq
Then, as first found in \cite{AMM, BE04} Tutte's recursion implies that generating functions of maps satisfy a topological recursion:
\bt[Topological recursion for numbers of maps]
\bea
\omega_{g,n}(z_1,z_2,\dots,z_n)
&=& \Res_{z\to \pm 1} K(z_1,z)\,\,\,\Big[ \omega_{g-1,n+1}(z,1/z,z_2,\dots,z_n) \cr
&& +\sum_{I\uplus I'=\{z_2,\dots,z_n\}; h+h'=g}' \omega_{h,1+\# I}(z,I)\,\omega_{h',1+\# I'}(1/z,I') \Big] \cr
\eea
where $\sum'$ means that we exclude from the sum the two cases $(I=\{z_2,\dots,z_n\},h=g)$ and $I=(\emptyset,h=0)$, and where the recursion kernel $K$ is:
\beq
K(z_1,z) = \frac{\,dz_1}{2\gamma}\,\frac{\frac{1}{z_1-z}-\frac{1}{z_1-1/z}}{(y(z)-y(1/z))}\,\frac{1}{(1-z^{-2})\,dz}
\eeq

\et

This recursion is very efficient at computing, for instance it easily gives:
\bea
\omega_{0,3}(z_1,z_2,z_3)
&=&  \frac{-1}{2\gamma y'(1) }\,\,\frac{1}{(z_1-1)^2}\,\frac{1}{(z_2-1)^2}\,\frac{1}{(z_3-1)^2} \cr
&& + \frac{1}{2\gamma y'(-1) }\,\,\frac{1}{(z_1+1)^2}\,\frac{1}{(z_2+1)^2}\,\frac{1}{(z_3+1)^2} .
\eea
For quadrangulations that gives
$$
\omega_{0,3}(z_1,z_2,z_3)
=  \frac{1}{4t-2\gamma^2}\,\left(\,\frac{1}{\prod_{i=1}^3 (z_i-1)^2}-\frac{1}{\prod_{i=1}^3 (z_i+1)^2} \right).
$$

%\bea
%\om_{1,1}(z_0)
%&=& {-1\over 16 \gamma y'(1) }\left( {1\over (z_0-1)^4} + {1\over (z_0-1)^3}  -\, {1 + {y''(1)\over y'(1)} + {y'''(1)\over 3y'(1)}\over 2 (z_0-1)^2} \right) \cr
%&& + {1\over 16 \gamma y'(-1) }\left( {1\over (z_0+1)^4} - {1\over (z_0+1)^3}  -\, {1 - {y''(-1)\over y'(-1)} + {y'''(-1)\over 3y'(-1)}\over 2 (z_0+1)^2} \right) \cr
%\eea
%
For genus $1$, it gives for quadrangulations
\beq
\om_{1,1}(z)
= -z\,\, \frac{t_4\gamma^4 z^4 + (t-5 t_4\gamma^4) z^2 + t_4\gamma^4}{(t-3t_4\gamma^4)^2\,\,(z^2-1)^4}
\eeq
which implies that the number of rooted quadrangulations of genus one, with $n$ faces (including the marked face) is:
\beq\label{nbquadrg1}
\#\,\,{\rm genus}\,1\, , \,n\,{\rm faces}:\qquad \quad
\frac{3^{n}}{6}\,\left(\frac{(2n)!}{n!\,n!}-2^n\right)
%= \frac{(12)^n}{6}\,\left(\begin{pmatrix} n-1/2 \cr n \end{pmatrix} - 1 \right)
\eeq
Similarly, the recursion easily gives $\omega_{2,1}$, but the result is too big to be written here, we shall only give the result that the number of rooted quadrangulations of genus 2, with $n+2$ faces (including the marked face) is:
\beq\label{nbquadrg2}
\#\,\,{\rm genus}\,2\, , \,n+2\,{\rm faces}:\qquad \quad
\frac{(12)^n}{2}\,\left(14\,\begin{pmatrix} n+5/2 \cr n \end{pmatrix} - 13 \,\begin{pmatrix} n+2 \cr n \end{pmatrix} - \begin{pmatrix} n+3/2 \cr n \end{pmatrix}
\right)
\eeq
%\beq
%\frac{3^{n-2}}{2}\,\left(\frac{7}{30}\,\frac{(2n+1)!}{(n-2)!\,n!}-\frac{(2n-3)!}{(n-2)!\,(n-2)!}-\frac{13}{2}\,2^{2n-4}\,n(n-1)  \right)
%\eeq
And for genus $3$, the topological recursion gives
\bea\label{nbquadrg3}
\#\,\,{\rm genus}\,3\, , \,n+4\,{\rm faces}:\qquad
(12)^n\,\Big(-2450\,\begin{pmatrix} n+5 \cr n \end{pmatrix}
+3033\,\begin{pmatrix} n+9/2 \cr n \end{pmatrix} \cr
-291\,\begin{pmatrix} n+4 \cr n \end{pmatrix}
+292\,\begin{pmatrix} n+7/2 \cr n \end{pmatrix}
\Big)
\eea

Again, we see that the "topological recursion" is an efficient method to effectively compute numbers of maps of any genus.

\section{How it arose}

The purpose of the present section is to recall how the topological recursion (which is mostly a geometric notion) was initially discovered from the study of large random matrices, and then happened to have a much broader reach in a geometric setting.

%\subsection{Random matrices and geometry of curves}

%\begin{enumerate}

\subsection{Introductory remark: Invariants in Geometry\label{secintroinv}}

In geometry, people are interested in finding "invariants", that is a collection of "numbers" (often integer or rational numbers) associated to a geometrical object, for instance to an algebraic curve.
Most often, those numbers are gathered into coefficients of a polynomial, or of a formal series, so by extension, we shall consider that invariants are formal series, typically $\in \mathbb Q[[\mathbf t]]$ for some set of formal parameters $\mathbf t=\{t_i\}$.

In enumerative geometry, people are interested in enumerating some objects within a geometry, for instance Gromov-Witten invariants "count the number of holomorphic embeddings of Riemann surfaces into a geometric space (for instance a Calabi-Yau manifold)".

Many questions may arise:

How these invariants get deformed under deformations of the geometry ? Do they satisfy nice differential equations (for instance $\partial/\partial t_i$) ?

What happens near singularities ?

Do we have modular properties (it is observed that many of the series appearing in enumerative geometry are actually modular forms) ?

Can different geometries have the same invariants ? In other words can the invariants discriminate the geometries, for instance in knot theory, Jones polynomials or HOMFLY polynomials were introduced in order to address that question.

And before all, how to compute those numbers in practice ?
Is there a universal method to do it ?

\subsection{Random Matrices}

A surprising answer came from a seemingly totally unrelated question in probabilities: what is the large size asymptotic statistics of a random matrix  ?

In random matrices, one is interested in the statistical properties of the spectrum, especially in the large size limit.
The density of eigenvalues converges (in most cases) towards a continuous density function, often called the "{\bf equilibrium measure}".
Very often (with reasonable choices of a random matrix probability law), the equilibrium measure is found to have a compact support (not necessarily connected), and happens to be an algebraic function.
This means that there is an algebraic curve related to the random matrix model.

For example, the equilibrium measure for eigenvalues of a Gaussian random matrix, is the famous "Wigner's semi-circle"
$$
\rho(x)dx = \frac{1}{2\pi}\,\sqrt{4-x^2}\,\,\,\,\mathbf{1}_{[-2,2]}\,\,dx
$$
it is described by the algebraic curve $y^2 = x^2-4$, where $y=2\ii\pi\rho(x)$ is the equilibrium density, supported on the segment $[-2,2]$.

Another famous example is "Marchenko--Pastur law", for singular values of an $M\times N$ random Gaussian matrix with variance $\sigma^2$
$$
\rho(x)dx = \frac{N}{2\pi\,M\sigma^2}\,\frac{\sqrt{\frac{M}{N}\sigma^4 - \frac{M^2}{N^2}\sigma^4-(x-\sigma^2)^2}}{x}\,\,\mathbf{1}_{[a,b]}\,\,dx
$$
(where $a$ and $b$ are the zeroes of the square--root term).
It is also algebraic.

\medskip

\subsubsection{Large size expansions}

Around 2004 it was observed \cite{AMM, BE04,CE06, ChEynbeta} that the knowledge of the equilibrium measure, is sufficient to recover the asymptotic expansion of every expectation value, and {\bf to  all orders} in the asymptotic expansion !

In other words, if ${\cal S}$ is the plane algebraic curve of the equilibrium measure, then all correlation functions are obtained as universal functionals of ${\cal S}$ only.

\smallskip
For example, a particularly interesting quantity is the "partition function".
Let $d\mu(M)$ be a (family depending on $N$ of) un--normalized measure on the set of Hermitian matrices of size $N$, the partition function is defined as
$$
Z = \int_{H_N} d\mu(M).
$$
Under "good assumptions" on the measure $d\mu$, the partition function has a large $N$ asymptotic expansion of the form
$$
\ln Z\sim \sum_{g=0}^\infty N^{2-2g}\,F_g.
$$
A main question in random matrix theory, is to compute the coefficients $F_g$ ?

In \cite{CE06,CEO06, EOFg} it was discovered that there exists a universal functional ${\cal F}_g:{\cal S}\mapsto {\cal F}_g({\cal S})$, such that, for many classes of random matrices:
$$
F_g = {\cal F}_g({\cal S}).
$$
The functional ${\cal F}_g:{\cal S}\mapsto {\cal F}_g({\cal S})$ is defined only in terms of the Riemannian geometry of the curve ${\cal S}$, it is often called "{\bf topological recursion}":
$$
\stackrel{{\rm spectral\,curve}}{\cal S} \quad \stackrel{{\rm topological\,recursion}}{\longrightarrow} \quad {\cal F}_g({\cal S}).
$$
For example, ${\cal F}_1({\cal S})$ is (up to a few extra factors beyond the scope of this introductory overview, see \cite{EOFg}) the log of the determinant of a canonical Laplacian on ${\cal S}$.

It was also discovered in \cite{AMM, BE04, EOFg} that there are also universal functionals $\omega_{g,n}:{\cal S} \mapsto \omega_{g,n}({\cal S})$ which compute the $g^{\rm th}$ order in the large $N$ expansion of the joint probability of $n-$eigenvalues (more precisely the cumulants of correlations of $n$ resolvents)
$$
\mathbb E\left(\prod_{i=1}^n \Tr(x_i-M)^{-1}\right)_{\rm cumulant} \qquad
\mathop{\sim}_{N\to\infty} \qquad \sum_{g=0}^\infty N^{2-2g-n}\,\omega_{g,n}.
$$

Therefore, for random matrices, there exist some functionals $\omega_{g,n}$ (and we call ${\cal F}_g=\omega_{g,0}$) which compute all correlation functions from the geometry of the spectral curve alone. The functionals $\omega_{g,n}$ are defined by a recursion on $g$ and $n$ (we postpone the explicit writing of this recursion to section \ref{secdeftoprec} below, because it involves some substantial background of Riemannian geometry), or more precisely a recursion on $(2g+n-2)$. This is called the {\bf topological recursion}:
$$
\omega_{g,n} = {\rm computed\,\,from}\,\, \omega_{g',n'}\,\,{\rm with}\,\,2g'+n'-2<2g+n-2.
$$

Eventually, this means that the knowledge of ${\cal S}$ (which is the large $N$ equilibrium density of eigenvalues) allows to recover all correlation functions of the random matrix law, i.e. recover the random matrix probability law itself:
%$$
%\stackrel{{\rm probability\,law}}{d\mu(M)} \quad \stackrel{{\rm large}\,N}{\longrightarrow} \quad \stackrel{{\rm spectral\,curve}}{{\cal S}=\{x\mapsto y\}} \quad \stackrel{{\rm topological\,recursion}}{\longrightarrow} \quad \omega_{g,n}({\cal S})
%$$
%and vice versa
$$
\stackrel{{\rm spectral\,curve}}{{\cal S}}
 \quad \stackrel{{\rm topological\,recursion}}{\longrightarrow} \quad \omega_{g,n}({\cal S})
\quad {\longrightarrow}  \quad
\stackrel{{\rm probability\,law}}{d\mu(M)} .
$$

\medskip
\subsubsection{How to use random matrices for geometry ?}

Since the functionals $\omega_{g,n}$ which give expectation values are universal and do not require anything but the curve ${\cal S}$, one may try to apply these functionals $\omega_{g,n}$ to any arbitrary algebraic curve, independently of whether that algebraic curve ${\cal S}$ was related to a random matrix law or not. This is the idea proposed in \cite{EOFg}.

\smallskip
In some sense, {\bf the topological recursion defines a "pseudo-random matrix law" associated to any plane curve ${\cal S}$.}
$$
\stackrel{{\rm plane\,curve}}{{\cal S}} \quad \stackrel{{\rm topological\,recursion}}{\longrightarrow} \quad \omega_{g,n}({\cal S})
\quad {\longrightarrow}  \quad
\stackrel{{\rm pseudo\, probability\,law}}{d\mu(M)} .
$$

\smallskip

Since expectation values or correlations are numbers, we get a collection of functionals, which associate numbers to a curve ${\cal S}$.
As we mentioned in the introduction subsection \ref{secintroinv}, this defines  {\bf "invariants of a curve"}.
$$
\stackrel{{\rm curve}}{{\cal S}} \quad \stackrel{{\rm topological\,recursion}}{\longrightarrow} \quad \omega_{g,n}({\cal S})
={\rm invariants\,of\,}\,{\cal S}.
$$

We thus have a definition of a family $\omega_{g,n}({\cal S})$ of invariants of a plane curve ${\cal S}$.
We shall call them the {\bf "symplectic invariants of ${\cal S}$"} or the {\bf "TR (topological recursion) invariants"} of ${\cal S}$.

\smallskip

{\em {\bf remark}: Let us emphasize some points here: not all algebraic curves can come from probabilities of random matrices, because probabilities have some real and positivity properties.
However, since the functional relations are analytical, they also apply to curves which don't have any positivity properties.

For example, the function $y=\frac{1}{4\pi}\,\sin{2\pi\sqrt{x}}$ appearing as the spectral curve for the Weil-Petersson volumes, can never be the density of eigenvalues of a random matrix.}

{\em {\bf remark}: The topological recursion, which associates  $\omega_{g,n}$  to a spectral curve ${\cal S}$, is defined only in terms of the Riemannian geometry on ${\cal S}$.
As we shall see, the topological recursion only uses local properties of the curve (residues), and thus it extends analytically to curves which are not necessarily algebraic.
}

\medskip
\subsection{Link with Mirror symmetry}

The topological recursion (TR) thus associates invariants to a spectral curve.
And we see that there are many examples of enumerative geometry, whose solution coincides with the TR invariants of a spectral curve appearing naturally in the problem.

\medskip

$\bullet$ {\bf A-model}. Enumerative geometry problems are often called "A-model". They deal with measuring "volumes" of moduli spaces (possibly discrete), i.e. counting configurations.
These moduli spaces are often equipped with a real symplectic structure.
Moduli spaces ${\cal M}$ are often of infinite dimension, equipped with some gradings, such that moduli spaces of given degree ${\cal M}(\mathbf d)$ are finite dimensional.
Thus the counting problem involves some formal variables $\mathbf t = (t_1,t_2,t_3,\dots)$ in order to define formal generating series:
$$
W(\mathbf t) = \sum_{{\mathbf d}={\rm degrees}} \,\,{\rm Volume}({\cal M}({\mathbf d}))\,\,\,\prod_i t_i^{d_i}
$$
Depending on the context, those parameters $t_i$ may be called "fugacities", "K\"ahler parameters", "Boltzman weights", "coupling constants", "spectral parameters",...
The generating series $W(\mathbf t)$ are often called "amplitudes" or "potential".

One is often interested in studying how the amplitudes depend on the parameters, thus unravelling the symplectic structure of the moduli space.

$\bullet$ {\bf B-model}. The equilibrium spectral curve of a pseudo-random matrix (or more precisely its Stieljes transform), is an analytical function $y=f(x)$, defined on a Riemann surface with a complex structure.
The pseudo-random matrix probability law, an thus the spectral curve, may depend on some parameters $\hat{\mathbf t}=(\hat t_1,\hat t_2,\dots)$.
Deformations of the probability law, induce deformations of the spectral curve, and in particular deformations of its complex structure.

B-model Amplitudes are the expectation values of the pseudo-random matrix, they can be computed as a universal functional of the spectral curve,  by the "topological recursion":
$$
{\rm Spectral\,curve}\,\,{\cal S}(\hat{\mathbf t})
\quad \stackrel{{\rm topological\,recursion}}{\longrightarrow} \quad {\rm amplitudes}\,\,\hat W(\hat{\mathbf t}) = \omega_{g,n}({\cal S}).
$$

One is often interested in studying how the amplitudes depend on the parameters, thus unravelling the complex structure of the moduli space.

\medskip

{\bf Mirror symmetry} is the claim that an A-model is dual to a B-model and vice versa, and there exists a "mirror map" $\mathbf t \mapsto \hat{\mathbf t}$ such that
$$
\hat W(\hat{\mathbf t}) = W(\mathbf t).
$$
One of the problems of mirror symmetry, is to identify which spectral curve should be associated to an enumerative geometry problem, i.e. find the mirror of the A-model geometry, and the mirror map for the parameters.

Most often the spectral curve happens to be a very simple and natural geometric object from the A-model point of view. It is typically the "most probable shape" of the objects counted in the A-model, in some "large size limit".
This is the case for random matrices, the spectral curve is the large size limit of the eigenvalue density.
Another example occurs in counting plane partitions, where the spectral curve is the shape of the limiting plane partition (often called arctic circle).
But there is unfortunately no general recipe of how to find the spectral curve mirror of a given A-model.

Many examples have been proved, many others are conjectures.
Here is a short non--exhaustive table of examples:

\framebox{
\begin{tabular}{l|l}

 A-model & B-model \cr

 moduli-space & Spectral curve \cr

\hrulefill & \hrulefill \cr

Kontsevich-Witten intersection numbers & $y^2=x$ \cr

$W_{g,n} = \sum_{\mathbf d} <\tau_{d_1}\dots\tau_{d_n}>_g \prod_{i=1}^n \frac{(2d_i-1)!!}{z_i^{2d_i+2}}\,dz_i$   & \cr

\hrulefill & \hrulefill \cr

Weil-Petersson volumes ${\cal V}_{g,n}(L_1,\dots,L_n)$ & $y=\frac{\sin{(2\pi\sqrt{x})}}{4\pi}$ \cr

\hrulefill & \hrulefill \cr

Hurwitz numbers  $H_g(\mu)$ & $y\,\ee{-y}=\ee{x}$ \cr

 \hrulefill & \hrulefill \cr

Random Matrix:  asymptotic expansion & $y=2\ii\pi \rho_{\rm eq}(x)$ \cr

of correlation functions  & \cr

$ \mathbb E(\prod_i \Tr (x_i-M)^{-1})$ & \cr

 \hrulefill & \hrulefill \cr

Toric Calabi-Yau  Gromov-Witten invariants & mirror curve $H(\ee{x},\ee{y})=0$ \cr

 \hrulefill & \hrulefill \cr

Knot theory  Jones polynomial & A-polynomial $A(\ee{x},\ee{y})=0$ \cr
& character variety \cr

\end{tabular}
}

%\end{enumerate}
(we detail some of those examples here below).

\medskip
\subsection{Some applications of symplectic invariants}

By definition, when ${\cal S}$ is the large $N$ spectral curve of a random matrix law, $\omega_{g,n}({\cal S})$ computes the $g^{\rm th}$ large $N$ order of the $n-$point correlation function  of resolvants:
$$
\mathbb E\left(\prod_{i=1}^n \Tr \frac{dx_i}{x_i-M}\right)_{\rm connected}
= \sum_{g=0}^\infty N^{2-2g-n}\,\omega_{g,n}.
$$

Is there other plane curves ${\cal S}$ for which those invariants compute something interesting ?

\medskip

The answer is YES: many classical geometric invariants, including Gromov-Witten invariants, or knot polynomials, can be obtained as the invariants of a plane curve ${\cal S}$ closely related to the geometry.

{\bf Examples:}

\smallskip

$\bullet$ Let $\mathfrak{X}$ be a local toric Calabi-Yau 3-fold \cite{Fulton93, BouchardToric}, and let ${\cal W}_{g,n}(\mathfrak{X})$ be the genus $g$ and $n$ boundary open Gromov--Witten invariant of $\mathfrak{X}$ (i.e. roughly speaking, the formal series whose coefficients count the number of holomorphic immersions of a genus $g$ Riemann surface with $n$ boundaries, such that the boundaries are mapped into a given Lagrangian submanifold \cite{KatzSheldon2001}). It is well known that the mirror \cite{mirrorbook} of $\mathfrak{X}$ is another Calabi-Yau 3-fold, of the form
$$
\{(x,y,u,v)\in \mathbb C^4 \, | \,\, H(\e^{x},\e^{y})=uv \}
$$
where $H$ is some polynomial found from the moment map of $\mathfrak{X}$.
This is an hyperbolic bundle over $\mathbb C^*\times \mathbb C^*$. The fibers are singular over the plane curve $H(\e^{x},\e^{y})=0$.
We call that plane curve  ${\cal S}=\hat{\mathfrak{X}}$, and we call it the mirror curve of $\mathfrak{X}$.

Then Mari\~no and co conjectured in \cite{Mar06,BKMP}, and it was proved in \cite{EOBKMP,LiuBKMP} that:
\bt[Topological recursion for toric CY 3folds] (called BKMP conjecture \cite{Mar06,BKMP}, first proved in \cite{EOBKMP}, and for CY orbifolds in \cite{LiuBKMP}).

The Gromov--Witten invariants ${\cal W}_{g,n}(\mathfrak{X})$ are the topological recursion invariants of ${\cal S}=\hat{\mathfrak{X}}$ the mirror curve of $\mathfrak{X}$:
$$
{\cal W}_{g,n}(\mathfrak{X}) = \omega_{g,n}(\hat{\mathfrak{X}}).
$$
\et

In fact, special cases of this theorem were first proved in \cite{ChenLin2009,ZhouJian2009,NS:2010,NS:2011}. The idea of the proof of \cite{EOBKMP,LiuBKMP}, is that the recursive structure of the topological recursion can be encoded as graphs (see def. \ref{defgraphs} below), and thus the $\omega_{g,n}(\hat{\mathfrak{X}})$ can be written as sums of weighted graphs. Those graphs, up to some combinatorial manipulations, happen to coincide with the localization graphs of Gromov-Witten invariants \cite{Diaco,LiuReview}. Thus it is mostly a combinatorial proof.

\medskip

$\bullet$ Another famous example (still conjectured) concerns knot polynomials.

Let $\mathfrak{K}$ be a knot embedded in the 3-dimensional sphere $S^3$.
The character variety of $\mathfrak{K}$ is the locus of eigenvalues of holonomies of a flat $SL(2)$ connection on the knot complement $S^3\setminus \mathfrak{K}$. This character variety is algebraic and defines an algebraic curve, called the $A$-polybomial of $\mathfrak{K}$:
$$
A(X,Y)=0.
$$
The colored-Jones polynomial $J_N(q)$, of color $N$, is defined as the Wilson loop \cite{Witten89} of a flat $SL(2)$ connection on $S^3\setminus \mathfrak{K}$, in the spin $N-1$ representation of $SL(2)$.
The Jones polynomial $J_N(q)\in\mathbb C[q]$ depends on $N$ (which labels the representation) and is a polynomial of a variable $q$.
Let us denote
$$
\hbar = \ln q \qquad , \qquad x = N\ln q.
$$
Then, it is conjectured \cite{DiFuji1, DiFuji1, GSq, GHSa, BEknot} that in the limit where $\hbar\to 0$ and $x=O(1)$, one has the asymptotic expansion:
$$
\ln J_N(q) \sim \sum_{k=-1}^\infty \hbar^{k}\,S_k(x)
\qquad , \quad
S_k(x)=
\sum_{2g-2+n=k} \frac{1}{n!} \int^x\dots\int^x \omega_{g,n}({\cal S})
$$
where ${\cal S}$ is the character variety of $\mathfrak{K}$, of equation $A(\ee{x},\ee{y})=0$.
(the more precise statement can be found in \cite{BEknot}).
In other words:

{\bf Conjecture: The Jones polynomial of a knot, is a series in $\hbar$ whose coefficients are the principal symplectic invariants of its A-polynomial}.

If true (which is of course expected), this conjecture would be an extension of the famous "Volume conjecture" \cite{Kash96,GuMu06,Hikami}, and would imply a new understanding of what Jones polynomials are, in particular that Jones polynomials are Tau-functions of some integrable systems \cite{DiFuji1, DiFuji1, GSq, GHSa, BEknot}.

\section{The definitions of topological recursion and symplectic invariants\label{secdeftoprec}}

\subsection{Spectral curves}

The topological recursion associates invariants $\omega_{g,n}$ to a spectral curve.

Let us give some abstract definitions of a spectral curve.

There exists many definitions of what a spectral curve is, they are more or less equivalent, but formulated in rather different languages.

Let us adopt the following definition here, close to the one in \cite{EOFg}

\bd[Spectral curve]\label{defspcurve1}
A spectral curve ${\cal S}=({\cal C},x,y,B)$ is:

- a Riemann surface ${\cal C}$ not necessarily compact nor connected,

- a meromorphic function $x:{\cal C}\to \mathbb C$, The zeroes of $dx$ are called the branchpoints. We assume that there is a finite number of them on ${\cal C}$.

- the germ of a meromorphic function at each branchpoint. We denote it collectively $y$. In other words near a branchpoint $a$ of order $r_a$ we write
\beq
y=\{\td t_{a,k}\}_{a\in {\rm branchpoints},\, k\in \mathbb N}
\quad \Leftrightarrow \quad
y(p) \mathop{\sim}_{p\to a} \sum_{k=0}^\infty \td t_{a,k} (x(p)-x(a))^{k/r_a}
\eeq

- a symmetric 1-1 form $B$ on ${\cal C}\times {\cal C}$, having a double pole on the diagonal and analytical elsewhere, normalized such that, with any local parameter:
\beq
B(p,q) \mathop{{\sim}}_{p\to q} \frac{dz(p)\otimes dz(q)}{(z(p)-z(q))^2} + {\rm analytical\,at}\,\,q
\eeq
again, in fact all what is needed is that $B$ is the germ of some analytical function at the branchpoints.

\ed

\br
Since the topological recursion computes residues, in fact all what is needed to run the recursion, is "formal neighbourhoods of branchpoints", with $y$ and $B$ to be germs of analytical functions.

However, in most practical examples, the neighbourhoods of branchpoints form an actual Riemann surface, on which $y$ and $B$ are globally analytical, and the geometric structure of that Riemann surface impacts a lot the properties enjoyed by the invariants.
In other words, the invariants are always well defined, but they enjoy more properties if in addition the Riemann surface has structure, for instance  if it is connected and/or compact, and for instance if $B$ is globally meromorphic.

\er

Since all what is needed are germs of analytical functions at the branchpoints,
we may define the spectral curve from the data of its Taylor (or Laurent) expansion coefficients, and thus propose another definition:

\bd[Spectral curve, bis]\label{defspcurve2}

A spectral curve ${\cal S}=(\{{\td t}_{a,k}\},\{\hat B_{a,k;b,j}\})$ is a collection of

$\bullet$ a set of "branchpoints" ${\mathbf a}=\{a_1,a_2,\dots,a_N\}$.

$\bullet$ a family of times $\td t_{a,k}$ for each $a\in \mathbf a$.
They are related to $y$ by $y(p) \mathop{\sim}_{p\to a} \sum_{k=0}^\infty \td t_{a,k} (x(p)-x(a))^{k/r_a}$.

$\bullet$ the times $\hat B_{a,k;b,j}$ for each $(a,b)\in \mathbf a\times \mathbf a$.
They are related to $B$ by
\beq
B(p,q) \mathop{\sim}_{p\to a,q\to b} \delta_{a,b} {\stackrel{\circ}{B}}_{a}(p,q) + \sum_{k,l} \hat B_{a,k;b,l} \zeta_a(p)^k\,\zeta_b(q)^l\,\,d\zeta_a(p)\,d\zeta_b(q)
\eeq
where $\zeta_a(p) = (x(p)-x(a))^{1/r_a}$, and
\beq
{\stackrel{\circ}{B}}_{a}(p,q) = \frac{d\zeta_a(p)\,d\zeta_a(q)}{(\zeta_a(p)-\zeta_a(q))^2}
\eeq

\ed

We shall propose 3 equivalent definitions

\subsection{Definition by recursion (B-model side)}

For simplicity in this definition below, we assume all branchpoints to be simple, i.e. $r_a=2$, the general case is done in \cite{Prats:2010, BHLMR, BoEy:2012}.
We define $\sigma_a: U_a\to U_a$ the involution in a small neighbourhood $U_a$ of $a$, that exchanges the two sheets of $x^{-1}$ that meet at $a$, i.e. such that
$$x\circ \sigma_a = x.$$
$\sigma_a$ is called the {\bf local Galois involution} of $x$ (it permutes the roots of $x(p)-{\rm x}$).
\bd
We define by recursion on $\chi=2g+n-2$, the following forms on ${\cal C}^n$:

\beq
\omega_{0,1}(p) = y(p) dx(p)
\eeq
\beq
\omega_{0,2}(p,q)=B(p,q)
\eeq
and for $2g+n-2\geq 0$:
\bea
\omega_{g,n+1}(p_1,\dots,p_{n+1})
&=& \sum_{a\in{\rm branchpoints}} \Res_{q\to a}
K_a(p_1,q)\,\Big[ \omega_{g-1,n+2}(q,\sigma_a(q),p_2,\dots,p_{n+1}) \cr
&& + \sum'_{h+h'=g,\, I\uplus I'=\{p_2,\dots,p_{n+1}\}}
\omega_{h,1+\# I}(q,I)\,\omega_{h',1+\# I'}(\sigma_a(q),I') \Big] \cr
\eea

\beq
{\rm with\,the\,recursion\,kernel}\qquad
K_a(p_1,q) = \frac{-1}{2}\,\,\frac{\int_{\sigma_a(q)}^{q} \omega_{0,2}(p_1,.)}{\omega_{0,1}(q)-\omega_{0,1}(\sigma_a(q))}
\eeq

\ed

\br
It is not obvious from the definition, but an important property (which can be proved by recursion, see \cite{EOFg}) is that $\omega_{g,n}$ is always a symmetric $n$-form on ${\cal C}^n$. The definition gives a special role to $p_1$, but the result of the sum of residues is in fact symmetric in all $p_i$'s, this can be proved by recursion \cite{EOFg}.
\er

\br
When the branchpoints are not simple, if $r_a>2$, the general definition can be found in \cite{Prats:2010, BHLMR, BoEy:2012}.
In fact, branchpoints of higher order $r_a>2$ can be obtained by taking a limit of several simple branchpoints merging smoothly. It was proved in \cite{BoEy:2012} that the limit of the definition with simple branchpoints, indeed converges to that of \cite{Prats:2010, BHLMR, BoEy:2012}.
In other words, higher order branchpoints, can be recovered from simple branchpoints.
This is why, for simplicity, we shall focus on simple branchpoints here.

For specialists, higher order branchpoints correspond to not necessarily semi--simple Frobenius manifold structures in Givental's formalism.
\er

Examples of applications of the definition for $(g,n)=(0,3)$:
\beq\label{W03gen}
\omega_{0,3}(p_1,p_2,p_3)
= \sum_a \Res_{q\to a} K_a(p_1,q) \,\Big[ B(q,p_2) B(\sigma_a(q),p_2) + B(q,p_1) B(\sigma_a(q),p_1) \Big]
\eeq
\beq\label{W11gen}
{\rm and\, for\,}\,(g,n)=(1,1)
\qquad \quad
\omega_{1,1}(p_1)
= \sum_a \Res_{q\to a} K_a(p_1,q) \,B(q,\sigma_a(q)).
\eeq

\bd\label{defFg}
When $n=0$ we define $\omega_{g,0}$ (denoted $F_g\equiv \omega_{g,0}$) by:
\beq
g\geq 2 \, , \qquad
F_g = \omega_{g,0} = \frac{1}{2-2g}\,\,\sum_a \Res_{q\to a} \,\,\omega_{g,1}(q)\,\Phi(q)
\eeq
where $d\Phi=\omega_{0,1}$ ($F_g$ is independent of a choice of integration constant for $\Phi$).

The definition of $F_1$ and $F_0$ is given in \cite{EOFg}, but we shall not write it in this short review.

\ed

\subsection{Definition as graphs}

The recursive definition above can conveniently be written in a graphical way.

For example expression \eqref{W03gen} or \eqref{W11gen} are easily written in terms of graphs:

$\bullet$ associate to each $B(p,q)$ factor, a non-oriented line from $p$ to $q$,

$\bullet$ associate to each $K_a(p,q)$ factor, an oriented line from $p$ to $q$, whose end $q$ has a "color" $a$,

$\bullet$ associate to each Residue $\Res_{q\to a}$  a tri-valent planar vertex of "color" $a$, with one ingoing edge (it must be oriented) and two outgoing edges (not necessarily oriented) the left one labeled with the point $q$ and the right one labeled with the point $\sigma_a(q)$.

$\bullet$ The value of a graph is then obtained by computing residues at the vertices of the product of $B$'s and $K$'s of edges.

For example \eqref{W03gen}
$$
\omega_{0,3}(p_1,p_2,p_3)
= \sum_a \Res_{q\to a} K_a(p_1,q) \,\Big[ B(q,p_2) B(\sigma_a(q),p_2) + B(q,p_1) B(\sigma_a(q),p_1) \Big]
$$
is represented by:
$$\includegraphics[scale=0.5]{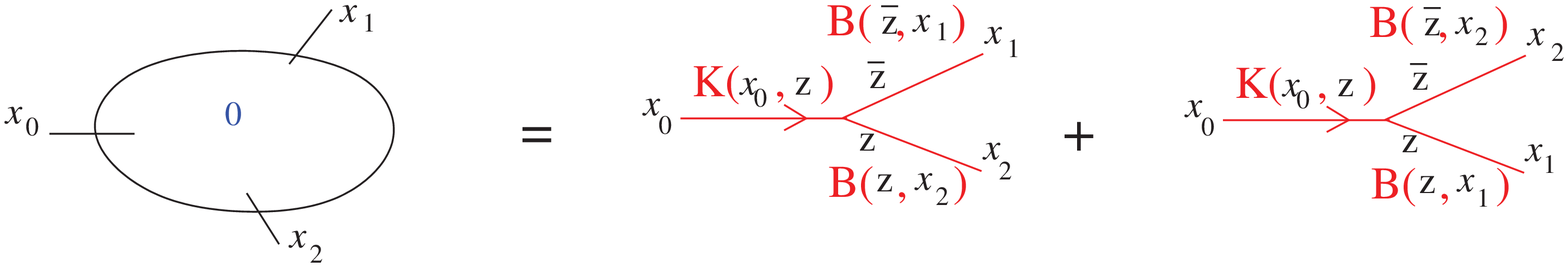}$$
$$
{\rm and\, }\,\qquad \qquad
\omega_{1,1}(p_1)
= \sum_a \Res_{q\to a} K_a(p_1,q) \,B(q,\sigma_a(q))
\qquad\qquad
{\rm is\,\,represented\,\,by}
$$
$$\includegraphics[scale=0.5]{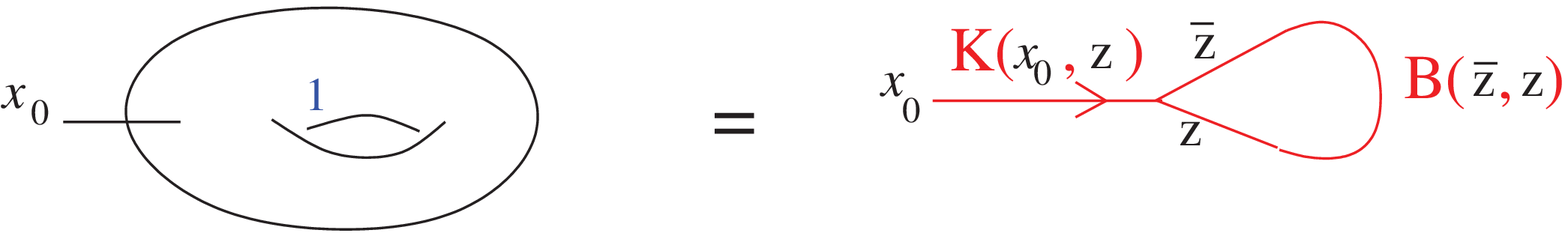}$$

\medskip

Therefore, following \cite{EOFg} we define the following set of graphs:

\bd\label{defgraphs}
For any $k\geq 0$ and $g\geq 0$ such that $k+2g-2>0$, we define:

Let ${\cal G}_{k+1}^{(g)}(p,p_1,\dots,p_k)$ be the set of connected trivalent graphs defined as follows:
\begin{enumerate}

\item there are $2g+k-1$ tri-valent vertices called vertices.
\item there is one 1-valent vertex labelled by $p$, called the root.
\item there are $k$ 1-valent vertices labelled with $p_1,\dots,p_k$ called the leaves.
\item There are $3g+2k-1$ edges.
\item Edges can be arrowed or non-arrowed. There are $k+g$ non-arrowed edges and $2g+k-1$ arrowed edges.
\item The edge starting at $p$ has an arrow leaving from the root $p$.
\item The $k$ edges ending at the leaves $p_1,\dots, p_k$ are non-arrowed.
\item The arrowed edges form a "spanning\footnote{It goes through all vertices.} planar\footnote{planar tree means that the left child and right child are not equivalent. The right child is marked by a black disk on the outgoing edge.} binary skeleton\footnote{a binary skeleton tree is a binary tree from which we have removed the leaves, i.e. a tree with vertices of valence 1, 2 or 3.} tree" with root $p$. The arrows are oriented from root towards leaves. In particular, this induces a partial ordering of all vertices.
\item There are $k$ non-arrowed edges going from a vertex to a leaf, and $g$ non arrowed edges joining two inner vertices. Two inner vertices can be connected by a non arrowed edge only if one is the parent of the other along the tree.
\item If an arrowed edge and a non-arrowed inner edge come out of a vertex, then the arrowed edge is the left child. This rule
only applies when the non-arrowed edge links this vertex to one of its descendants (not one of its parents).

\end{enumerate}

Then, we define the weight of a graph as:

\beq
w(G) = \prod_{v\in\{{\rm vertices}\}} \Res_{q_v\to a_v}
\,\, \prod_{e=(p,q)\in\{{\rm unarrowed\,edges}\}} B(p,q)
\,\, \prod_{e=(p\mapsto q)\in\{{\rm arrowed\,edges}\}} K_{a_p}(p,q)
\eeq
where the order of taking the residues is by following the arrows from leaves to root (deeper vertices are integrated first).

Then, the definition of $\omega_{g,n}({\cal S})$ is:
\beq
\omega_{g,n}(p_1,\dots,p_n) = \sum_{G\in {\cal G}_{g,n}(p_1,\dots,p_n)} w(G).
\eeq

\ed

Those graphs are merely a notation for the previous recursive definition they are merely a convenient mnemotechnic rewriting.

This graphical notation is very convenient, it is a good support for intuition and is very useful for proving some theorems.

The topological recursion can be graphically illustrated as follows:
$$\includegraphics[scale=0.6]{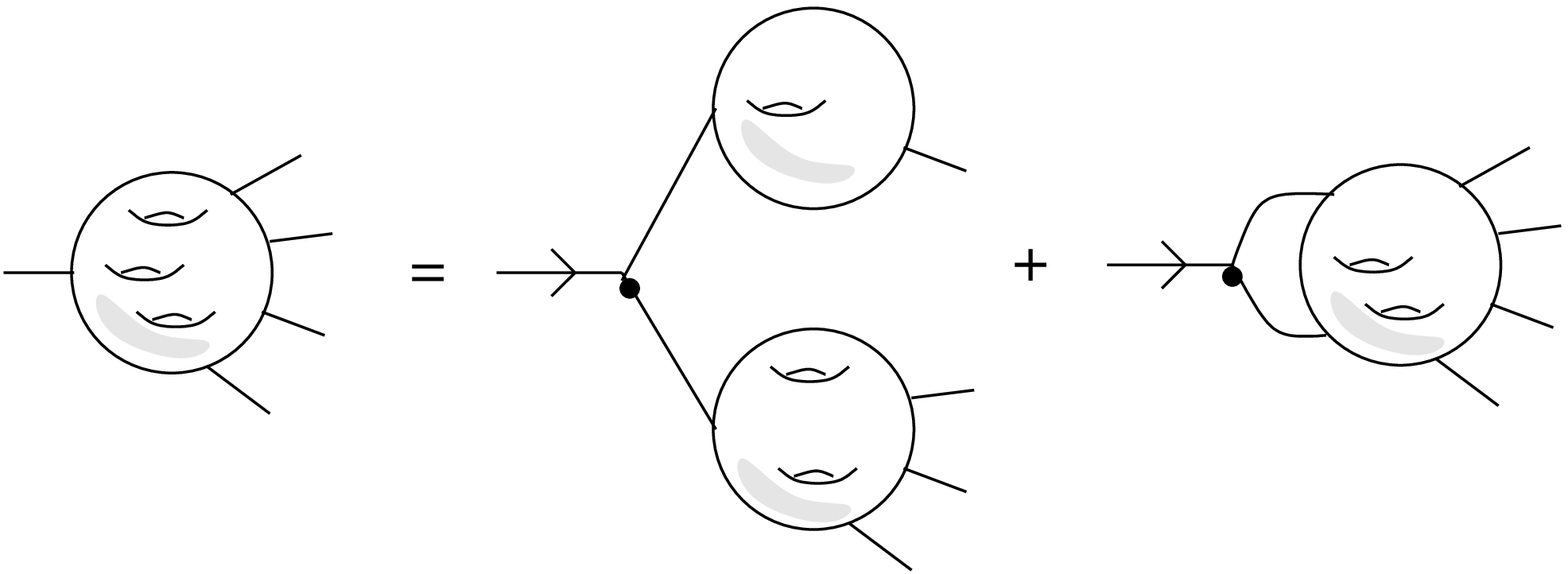}$$

\subsection{A-model side definition}

We propose another equivalent definition.

%This subsection requires some knowledge of algebraic geometry.
%For simplicity we shall again often assume that branchpoints are simple $r_a=2$, but what follows would also work for arbitrary $r_a$ with just more tedious notations.

Let ${\cal S}=({\cal C},x,y,B)$ a spectral curve, with branchpoints $\mathbf a=\{a\}$.
Near $a\in \mathbf a$, we define the local Laplace transforms
\bd[Laplace transforms]
\beq
\e^{-f_{(a,j)}(u)}=\frac{u^{3/2}\,\e^{u x(a)}}{2\,\sqrt{\pi}}\,\int_{\gamma_{(a,j)}}\,y\,dx\,\,\e^{-u x}
\eeq
where $\gamma_{(a,j)}$ is a "steepest descent path", i.e. in a neighbourhood $U_a$ of $a$ it is an arc included in $x^{-1}(x(a)+\mathbb R_+)$ (if $dx$ vanishes to order $r_a-1$ at $a$ ($x$  is locally $r_a:1$ at $a$), then there are $r_a-1$ such steepest descent paths, i.e. $j\in[1,\dots,r_a-1]$. For a simple branchpoint we have $r_a=2$, and there is only $j=1$, so we may drop the $j$ index).

Its large $u$ expansion doesn't depend on the neighbourhood $U_a$ and defines the "times":
\beq
f_{(a,j)}(u) \sim \sum_{k=0}^\infty t_{(a,j),k}\,u^{-k}.
\eeq
\ed

Similarly we Laplace transform $B$:
\bd
\beq
\hat B_{(a,j);(b,l)}(u,v) = \frac{\sqrt{uv}}{\pi}\int_{\gamma_{(a,j)}\times \gamma_{(b,l)}} \left(B(z_1,z_2)-\delta_{a,b}\,{\stackrel{\circ}{B}}_{a}(z_1,z_2)\right)\,\,\e^{-u (x(z_1)-x(a))}\,\e^{-v (x(z_2)-x(b))}
\eeq
where ${\stackrel{\circ}{B}}_{a}(z_1,z_2) = \frac{d\zeta_a(z_1)\,d\zeta_a(z_2)}{(\zeta_a(z_1)-\zeta_a(z_2))^2}$, with $\zeta_a(z)=(x(z)-x(a))^{1/r_a}$.

The large $u$ and $v$ expansion define the "times"
\beq
\hat B_{(a,j);(b,l)} \sim \sum_{m,n} \hat B_{(a,j),n;(b,l),m} u^{-n}\,v^{-m}.
\eeq

We shall also define the half Laplace transform
\beq
\check B_{(a,j)}(u,z) = \frac{\sqrt{u}}{\sqrt{\pi}}\int_{z'\in \gamma_{(a,j)}} B(z',z)\,\,\e^{-u (x(z')-x(a))}
\eeq
whose large $u$ expansion defines a basis of meromorphic 1-forms having a pole at $a$:
\beq
\check B_{(a,j)}(u,z) = \sum_k u^{-k}\, d\xi_{(a,j),k}(z)
\eeq

\ed

All this gives another definition of the notion of spectral curve:

\bd[Spectral curve, ter]\label{defspcurve3}

A spectral curve ${\cal S}=\{\,\{t_{\alpha,k}\},\{\hat B_{\alpha,k;\beta,l}\},\{d\xi_{\alpha,k}\} \}$ is the data of all the times.

\ed

This definition encodes in a slightly different way compared to def\ref{defspcurve1}, the Taylor expansions of all germs of analytical functions needed to run the recursion, which are much better encoded through Laplace transforms, as remarked in \cite{EynardMumford, DMSS,eynclasses1,eynclasses2,DOSS}.

We shall now use the spectral curve data to define a tautological cohomology class in the cohomological ring of some moduli space, and thus define an A-model potential.

First, we define the moduli space. Let us first assume that all branchpoints are simple, i.e. $r_a=2$ and thus the local Galois group is $\mathbb Z_2$:

\bd[Colored moduli space (simple branchpoints)]

Let
$$\mathbf a=\{a\}_{a={\rm branchpoints}}
\qquad , \quad N=\#\mathbf a.$$

We start by defining the following moduli space (not compact):
\beq
{\cal M}_{g,n}(\mathbf a)=\{(\Sigma,p_1,\dots,p_n,s)\}
\eeq
where $\Sigma$ is a genus $g$ nodal surface with $n$ smooth marked points $p_1,\dots,p_n$, and $s:\Sigma\to \mathbf a$ be a map constant in each component of $\Sigma$.

In fact ${\cal M}_{g,n}(\mathbf a)$ is merely a convenient notation for a union of smaller moduli spaces:
\beq
{\cal M}_{g,n}(\mathbf a) = \cup_{G = {\rm dual\,graphs}, N\,{\rm colored}} \prod_{v\in{\rm vertices}} \overline{\cal M}_{g_v,n_v}^{(a_v)}
\eeq
where $\overline{\cal M}_{g,n}^{(a)}$ are $N$ copies of $\overline{\cal M}_{g,n}$ labeled by the branchpoints $a$. The graphs $G$ are dual graphs of stable nodal surfaces, of total genus $g$ and $n$ smooth marked points. Vertices $v$ of $G$ carry a genus $g_v$, a number of marked or nodal points $n_v$, and a color $s_v\in \mathbf a$. We must have:
$$\forall\,v,\,\, 2-2g_v-n_v<0
\qquad , \quad {\rm and}\quad \sum_{v\in {\rm vertices\,of}\,G} (2-2g_v-n_v)=2-2g-n.$$

\ed

In fact this definition can be extended to multiple branchpoints $r_a>2$, with a local Galois group $\mathfrak G_a$ (most often $\mathbb Z_{r_a}$).

\bd[Colored moduli space (multiple branchpoints)]

Let
$$\mathbf a=\{\alpha\}=\{(a,j)\}_{a={\rm branchpoints}, j\in \mathfrak G_a}
\qquad , \quad N=\#\mathbf a.$$

We start by defining the following moduli space (not compact):
\beq
{\cal M}_{g,n}(\mathbf a)=\{(\Sigma,p_1,\dots,p_n,s)\}
\eeq
where $\Sigma$ is a genus $g$ nodal surface with $n$ smooth marked points $p_1,\dots,p_n$, and $s:\Sigma\to \mathbf a$ be a map constant in each component of $\Sigma$.

In fact ${\cal M}_{g,n}(\mathbf a)$ is merely a convenient notation for a union of smaller moduli spaces:
\beq
{\cal M}_{g,n}(\mathbf a) = \cup_{G = {\rm dual\,graphs}, N\,{\rm colored}} \prod_{v\in{\rm vertices}} B{\mathfrak G_{a_v}}\overline{\cal M}_{g_v,n_v}^{(a_v)}
\eeq
where $\overline{\cal M}_{g,n}^{(a)}$ are $\#\{a\}$ copies of $\overline{\cal M}_{g,n}$ labeled by the branchpoints $a$, and $B{\mathfrak G_{a}}$ the classifying space of the local Galois group ${\mathfrak G_{a}}$ at the branchpoint $a$, and $B{\mathfrak G_{a}}\overline{\cal M}_{g,n}$ is defined by the Chen-Ruan cohomology of $\mathbb C^3/\mathfrak G_a$. The graphs $G$ are dual graphs of nodal surfaces, of total genus $g$ and $n$ smooth marked points. Vertices $v$ of $G$ carry a genus $g_v$, a number of marked or nodal points $n_v$, and a color $s_v\in \mathbf a$.

\ed

\medskip

Then, we define the following tautological classes in the cohomological ring of the moduli space ${\cal M}_{g,n}(\mathbf a)$. We do the case of simple branchpoints for simplicity:

\bd[Tautological class of a spectral curve\label{defLambdaS}]

\beq
\Lambda({\cal S})
= \e^{\sum_k t_{s_*,k} \kappa_{k} + \frac{1}{2}\sum_{k,l} \hat B_{s_*,k;s_*,l} \sum_{\delta\in \partial{\cal M}_{g,n}(\mathbf a)} l_\delta*(\tau_{k}\tau_{l})}
\eeq
where

$\bullet$ $\tau_k=c_1(T^*_p)^k=\psi(p)^k$ is the $k^{\rm th}$ power of the 1st Chern class of the cotangent bundle at the marked or nodal point $p$ over $\overline{\cal M}^{(a_{s(p)})}_{g,n}$

$\bullet$ $\kappa_k$ is the $k^{\rm th}$ Mumford class \cite{Arbarello1996}, that is the pushforward $\pi*\psi(p_{n+1})^{k+1}$ of the $(k+1)^{\rm th}$ power of the 1st Chern class of the cotangent bundle at the $(n+1)^{\rm th}$ marked points in $p_{n+1}\in\overline{\cal M}_{g,n+1}$, under the forgetful map $\pi:\overline{\cal M}_{g,n+1}\to \overline{\cal M}_{g,n}$.

$\bullet$ $\partial {\cal M}_{g,n}(\mathbf a)$ is the set of boundary divisors of ${\cal M}_{g,n}(\mathbf a)$, or in other words it is the set of nodal points.
If $\delta\in \partial {\cal M}_{g,n}(\mathbf a)$, then $\delta$ is a nodal point, i.e. it is a pair of points $\delta=(p,p')$ with $p$ and $p'$ in two components (possibly the same) of $\Sigma$, corresponding to two vertices $v,v'$ (possibly the same) of the graph $G$.
$l_\delta*(\tau_{k}\tau_{l})$ denotes the class $\psi(p)^k \psi(p')^l$ pushed in $\overline{\cal M}_{g_v,n_v}^{(\sigma(p))}\times \overline{\cal M}_{g_{v'},n_{v'}}^{(\sigma(p'))}$.

\ed

Then the invariants  $\omega_{g,n}({\cal S})$ are given by:
\bt[A-model Invariants] (proved in \cite{eynclasses2}, see also \cite{DOSS}).\label{thomegaLambda}

\bea
\omega_{g,n}(z_1,\dots,z_n)
&=& \int_{{\cal M}_{g,n}(\mathbf a)} \Lambda({\cal S})\,\, \prod_{i=1}^n \check B_{s(p_i)}(1/\psi(p_i),z_i)  \cr
&=& \sum_{d_1,\dots,d_n} \int_{{\cal M}_{g,n}(\mathbf a)} \Lambda({\cal S})\,\, \prod_{i=1}^n \psi(p_i)^{d_i}\,\, d\xi_{s(p_i),d_i}(z_i)  \cr
\eea

\et

In fact this theorem and the definition of ${\cal M}_{g,n}(\mathbf a)$ means a sum over graphs of products at vertices of usual intersection numbers in some $\overline{\cal M}_{g_v,n_v}^{(a_v)}$'s, it is merely a short hand notation for the following sum:
\bea
\omega_{g,n}(z_1,\dots,z_n)
&=& 2^{3g-3+n}\sum_{{\rm graphs}\,G} \quad \sum_{\{d_h\}\in \mathbb Z^{\{{\rm half-edges}(G)\}}}\frac{2^{-\#{\rm edges}(G)}}{\#{\rm Aut}(G)} \cr
&& \prod_{e=(v,v')\in{\rm edges}(G)}  \hat B_{a_{v},d_{(v,e)}; a_{v'},d_{(v',e)}} \qquad \prod_{i=1}^n \,\, d\xi_{s(p_i),d_i}(z_i)  \cr
&&  \prod_{v\in{\rm vertices}(G)} \int_{\overline{\cal M}_{g_v,n_v}^{(a_v)}} \e^{\sum_k t_{a_v,k} \kappa_k}\quad  \prod_{h\in{\rm half-edges}(G)\,{\rm adjacent\,to}\,v} \psi(p_{h})^{d_{h}}\,\,\cr
\eea

This theorem is thus a mirror symmetry statement \cite{mirrorbook}. It was first proved in \cite{eynclasses1} for a single branchpoint, and then in \cite{eynclasses2} for the general case, and see also \cite{DOSS}.

{\bf Idea of the proof:}
Using the graphical definition def.\ref{defgraphs} of the $\omega_{g,n}$'s, by a recombination of vertices with the same colors, one finds that $\omega_{g,n}$ can be written as a sum over graphs of a Wick theorem  \cite{Kost, OrantinN.2008, Kostov2010}, where the edge weights are the $\hat B_{a,k;b,l}$'s, and it remains to compute the weights of vertices.

Since vertices are independent of the $\hat B_{a,k;b,l}$, they can be found from the case where all $\hat B_{a,k;b,l}$ vanish, and when there is only one branch point. This can be achieved by chosing the spectral curve ${\cal S}=(\mathbb CP^1, x:z\mapsto z^2,y:z\mapsto z,B(z,z')=\frac{dz dz'}{(z-z')^2})$, and shows that the weights of vertices \cite{EynardMumford, eynclasses1} are the Witten Kontsevich intersection numbers \cite{Konts, Witten}.
Therefore, this theorem is mostly of combinatorial nature. $\square$

\smallskip

\br
In fact this theorem is very similar to Givental's formalism \cite{givental-2001}. The only difference with Givental's formalism, is that it applies to more general situations. Givental's formalism applies to Gromov--Witten's theories, and thus applies only if the coefficients $\hat B$'s and $t$'s satisfy certain relationships which we don't assume here.
All this is explained in \cite{DOSS}.

\er

\medskip

{\bf Examples of applications of theorem \ref{thomegaLambda}:}

\medskip
$\bullet$ {\bf Weil-Petersson volumes:}

Chose  ${\cal S}=(\mathbb CP^1, x:z\mapsto z^2,y:z\mapsto\frac{1}{4\pi}\,\sin{(2\pi z)},B(z,z')=\frac{dz dz'}{(z-z')^2})$.
In that case, there is only one branchpoint at $z=0$. An easy computation yields $\hat B_{a,k;a,l}=0$, and the Laplace transform of $ydx$ yields:
\beq
\e^{-f(u)}
= \frac{u^{3/2}}{2\,\sqrt\pi}\,\int_{-\infty}^{\infty} \frac{\sin{(2\pi z)}}{4\pi} \,2zdz \,\,\e^{-u z^2}
= \frac{1}{4}\,\e^{-\pi^2/u}
\eeq
and we also find
\beq
d\xi_d(z) = \frac{(2d+1)!!}{2^d}\,\,\frac{dz}{z^{2d+2}}
\eeq
Definition \ref{defLambdaS} gives
$$
\Lambda({\cal S}) = \e^{\pi^2\kappa_1},
$$
and the theorem \ref{thomegaLambda} gives
\bea
\omega_{g,n}(z_1,\dots,z_n)
&=& 2^{5g-5+2n}\sum_{d_1,\dots,d_n} \,\prod_{i=1}^n d\xi_{d_i}(z_i)\,\,\,\int_{\overline{\cal M}_{g,n}} \e^{\pi^2\kappa_1}\,\,\prod_{i=1}^n \psi_i^{d_i}  \cr
&=& 2^{2g-2+n}\,\sum_{d_1,\dots,d_n} \,\prod_{i=1}^n \frac{(2d_i+1)!!\, dz_i}{z_i^{2d_i+2}}\,\,\,\int_{\overline{\cal M}_{g,n}} \e^{2\pi^2\kappa_1}\,\,\prod_{i=1}^n \psi_i^{d_i}  \cr
\eea
which are indeed the Weil-Petersson volumes of moduli spaces \cite{mulsaf, Xu, Mulase2006, EOWP}.
In other words, this theorem gives a very easy way to rederive Mirzakhani's recursion, by simply computing the Laplace transform of $y=\sin{(2\pi\sqrt{x})}/4\pi$.

\medskip
$\bullet$ {\bf Kontsevich-Witten:}

Chose  ${\cal S}=(\mathbb CP^1, x:z\mapsto z^2,y:z\mapsto z,B(z,z')=\frac{dz dz'}{(z-z')^2})$.
In that case, there is only one branchpoint at $z=0$. An easy computation yields $\hat B_{a,k;a,l}=0$, and the Laplace transform of $ydx$ yields:
\beq
\e^{-f(u)}
= \frac{u^{3/2}}{2\,\sqrt\pi}\,\int_{-\infty}^{\infty} z\,\, \,2zdz \,\,\e^{-u z^2}
= \frac{1}{2}
\eeq
and we also find
\beq
d\xi_d(z) = \frac{(2d+1)!!}{2^d}\,\,\frac{dz}{z^{2d+2}}
\eeq
Definition \ref{defLambdaS} gives
$$
\Lambda({\cal S}) = 2^{\kappa_0} = 2^{2g-2+n}.
$$
The theorem \ref{thomegaLambda} thus gives
\bea
\omega_{g,n}(z_1,\dots,z_n)
&=& 2^{2g-2+n}\,\sum_{d_1,\dots,d_n} \,\prod_{i=1}^n \frac{(2d_i+1)!!\, dz_i}{z_i^{2d_i+2}}\,\,\,\int_{\overline{\cal M}_{g,n}} \,\,\prod_{i=1}^n \psi_i^{d_i}  \cr
\eea
which are the Kontsevich-Witten \cite{Konts,Witten} intersection numbers \cite{EynardMumford, eynclasses1, mulsaf, Xu, Mulase2006}.

\medskip

$\bullet$ ELSV formula

Chose  ${\cal S}=(\mathbb CP^1, x:z\mapsto -z+\ln z, y:z\mapsto z, B(z,z')=\frac{dz dz'}{(z-z')^2})$.
Again there is only one branchpoint at $z=1$.
The Laplace transform of $ydx$ yields:
\bea
\e^{-f(u)}
&=& \frac{u^{3/2}\e^{-u}}{2\sqrt\pi}\,\int_{0}^{-\infty} z\,\frac{(1-z)\,dz}{z}\,\, z^{-u}\,\e^{uz} \cr
&=& \frac{\ii\,\sqrt{\pi}\,u^u \e^{-u}}{\,\sqrt{u}\,\Gamma(u)}  \cr
&=& \frac{\ii}{\sqrt{2}}\,\,\e^{-\sum_k \frac{B_{2k}}{2k(2k-1)}\,u^{1-2k}}
\eea
where $B_k$ are the Bernoulli numbers.
We leave the reader an exercise to compute the $\hat B_{a,k;a,l}$ and the $d\xi_{a,k}(z)$, and we just mention that:
\beq\label{MumfordLambda}
\Lambda_{\rm Hodge}=\e^{\sum_k \frac{B_{2k}}{2k(2k-1)}\,\left(\kappa_{2k-1}-\sum_i \psi_i^{2k-1} + \frac{1}{2}\sum_\delta \sum_{l=0}^{2k-2} (-1)^l\,\, l_\delta* \tau_{2k-2-l} \tau_l\right)}
\eeq
is the Hodge class \cite{mumford, eynclasses1}.
The theorem above easily gives the ELSV formula \cite{ELSV}, but we refer the reader to \cite{eynclasses1,eynclasses2} for a more detailed computation.

\medskip

$\bullet$ Mari\~no--Vafa formula \cite{MV01} for the topological vertex with framing $f$

Chose  ${\cal S}=(\mathbb CP^1, x:z\mapsto -f \ln z-\ln(1- z), y:z\mapsto \ln z, B(z,z')=\frac{dz dz'}{(z-z')^2})$.
Again there is only one branchpoint at $z=\frac{f}{f+1}$.
The Laplace transform of $ydx$ yields:
\bea
\e^{-f(u)}
&=& \frac{u^{1/2}\,((f+1)^{f+1}/f^f)^u}{2\sqrt\pi}\,\int_{0}^{1} (1-z)^u\,\,z^{f\,u}\,\frac{dz}{z} \cr
&=& \frac{u^{1/2}\,((f+1)^{f+1}/f^f)^u}{2(f+1)\sqrt\pi}\,\,\frac{\Gamma(f u) \Gamma(u)}{\Gamma((f+1)u)}  \cr
&=& \frac{1}{(f+1)\sqrt{2}}\,\,\e^{-\sum_k \frac{B_{2k}}{2k(2k-1)}\,u^{1-2k}\,\,(1+f^{1-2k}+(-1-f)^{1-2k})}
\eea
where $B_k$ are the Bernoulli numbers.
We leave the reader an exercise to compute the $\hat B_{a,k;a,l}$ and the $d\xi_{a,k}(z)$, and we just mention that using \eqref{MumfordLambda}, we see that the spectral curve's class here, is a product of 3 Hodge classes
\beq
\Lambda_{\rm Hodge}(1)\Lambda_{\rm Hodge}(f)\Lambda_{\rm Hodge}(-1-f)
\eeq
and the theorem above easily gives the Mari\~no--Vafa formula \cite{MV01,Liu2003,Liu2003a}, but we refer the reader to \cite{eynclasses1,eynclasses2} for a more detailed computation.

\section{Main properties}

Let us make a brief summary of some of the properties enjoyed by those invariants.

\medskip

$\bullet$ {\bf Symplectic invariance}

$\hat {\cal F}_g={\cal F}_g-\frac{1}{2-2g}\sum_\alpha (\Res_\alpha ydx)(\int_o^\alpha \omega_{g,1})$ is invariant under symplectomorphisms.
This means that if $\phi:\mathbb{C}P^1\times \mathbb{C}P^1 \to \mathbb{C}P^1\times \mathbb{C}P^1$ is a symplectomorphism (conserves $dx\wedge dy$), then
\bt
If ${\cal S}=({\cal C},x,y,B)$ is such that ${\cal C}$ is a compact Riemann surface, $x$ and $y$ are globally meromorphic functions on ${\cal C}$, and $B$ is the fundamental 2nd kind form \cite{BergSchif} on ${\cal C}\times {\cal C}$, normalized on a symplectic basis of cycles, then if $\phi$ is a symplectomorphism of $\mathbb{C}P^1\times \mathbb{C}P^1$ then
\beq
\hat {\cal F}_g(\phi*{\cal S}) = \hat{\cal F}_g({\cal S}).
\eeq
\et
This theorem \cite{EOFg, EOFgxy,EOFgxyer} is extremely powerful and useful. It allows to compare very easily some apparently unrelated enumerative problems, just by comparing their spectral curves. For instance it allows to find dualities.
A special case is $\phi:(x,y)\mapsto (y,-x)$, i.e.
\beq
\hat {\cal F}_g({\cal C},y,-x,B) = \hat{\cal F}_g({\cal C},x,y,B).
\eeq

Let us mention that the proof of that theorem is highly non--trivial, and it was proved so far only for algebraic spectral curves ($x$ and $y$ meromorphic on a compact ${\cal C}$), but it is believed to be valid in more general cases for example when $dx$ and $dy$ are meromorphic 1-forms on a compact ${\cal C}$, see \cite{BS:2011, BHLMR}.

\medskip
$\bullet$ {\bf Modular invariance}
Let  ${\cal S}=({\cal C},x,y,B)$ a spectral curve such that ${\cal C}$ is a compact Riemann surface of genus $\genus$,  and $B$ is the fundamental 2nd kind form on ${\cal C}\times {\cal C}$, normalized on a symplectic basis of cycles.
The modular group $Sp_{2\genus}(\mathbb Z)$ acts on $B$.
If  $\begin{pmatrix} a & b \cr c & d \end{pmatrix}\in Sp_{2\genus}(\mathbb Z)$, the period matrix is changed to $\tau \mapsto (d-\tau b)^{-1}\,(\tau a-c)$, and $B$ is changed to
\beq\label{Bmodular}
B(p,p')\mapsto B(p,p') + 2\pi\ii\,\, \sum_{i,j=1}^\genus \omega_i(p)\,(
b\,(d-\tau b)^{-1})_{i,j}\,\omega_j(p')
\eeq
where $\omega_i$ are the normalized holomorphic 1-forms on ${\cal C}$.
Then
\bt
${\cal F}_g({\cal C},x,t,B)$ is an almost modular form under the modular group $Sp_{2\genus}(\mathbb Z)$ acting on $B$.
\et
The proof of this theorem appeared in \cite{EOFg} and \cite{EOMmodular}, and follows easily from the graphical decomposition of def \ref{defgraphs}. Indeed, $B$ appears only in edges of the graphs, and eq.\eqref{Bmodular} amounts to cutting edges. The effect of a modular transformation thus produces dual graphs of degenerate Riemann surfaces, with factors of $b(d-\tau b)^{-1}$ at degeneracies, and thus coincides with the transformations of almost modular forms.

\medskip

$\bullet$ Deformations and Form cycle duality

The tangent space to the space of spectral curves at ${\cal S}=({\cal C},x,y,B)$ is the space of meromorphic forms on ${\cal C}$.
Lat us chose $B$ to be the fundamental 2nd kind differential on ${\cal C}$, see \cite{MumTata, fay, EOFg}. Then $B$ provides the kernel for a form--cycle duality pairing, namely the meromorphic form $\Omega$ dual to a cycle $\Omega^*$ is
\beq
\Omega(p) = \oint_{p'\in \Omega^*} B(p,p').
\eeq
(here we call cycle $\Omega^*$ any linear form on the space of meromorphic forms, i.e. an element of the dual of ${\cal M}^1({\cal C})$).

Then we have
\bt
Let $\Omega$ be a tangent vector to the space of spectral curves, i.e. a meromorphic 1-form on ${\cal C}$, and $\partial_\Omega$ be the derivative in the direction of $\Omega$, then we have
\beq
\partial_\Omega \,\omega_{g,n} = \oint_{\Omega^*} \omega_{g,n+1}
\eeq
\et

This theorem first proved in \cite{CE06, EOFg} follows easily from the graphical decomposition of def \ref{defgraphs}. Indeed, one just has to see how $\partial_\Omega$ acts on the kernels $K$ and $B$, i.e. on the edges of the graphs, and it produces exactly the graphs of $\omega_{g,n+1}$.

Special cases of that theorem are:
\beq
\partial_\Omega\, ydx = \partial_\Omega \,\omega_{0,1} = \oint_{\Omega^*} \omega_{0,2} = \oint_{\Omega^*} B = \Omega.
\eeq
we thus recover that $\Omega$ is the derivative of $\omega_{0,1}=ydx$ i.e. a meromorphic 1-form on $T^*{\cal C}$.
Another example is
\beq
\partial_\Omega\, {\cal F}_0 = \oint_{\Omega^*} \omega_{0,1} = \oint_{\Omega^*} ydx
\eeq
which means that ${\cal F}_0$ is the prepotential, this relation is Seiberg-Witten's duality.
Yet another example is
\beq
\partial_\Omega\, B(p_1,p_2) = \oint_{p_3\in\Omega^*} \omega_{0,3}(p_1,p_2,p_3)  = \sum_a \Res_{q\to a}  \frac{B(q,p_1)B(q,p_2)\,\Omega(q)}{dx(q)dy(q)}
\eeq
which is known as the Rauch variational formula for the fundamental 2nd kind form $B$.
Another example is:
\beq
\partial_\Omega\, {\cal F}_1 = \oint_{\Omega^*} \omega_{1,1}
\eeq
which means that ${\cal F}_1$ is (up to some details which we don't enter here) the Bergman Tau function of Kokotov--Korortkin \cite{KoKo2}.

Again, this theorem is very powerful.

\medskip

$\bullet$ {\bf Dilaton equation}

This is an equation saying that
\bt
For any $(g,n)$ such that $2g-2+n>0$ we have
\beq
\sum_a \Res_{q\to a}\,\, \omega_{g,n+1}(p_1,\dots,p_n,q)\,\Phi(q) = (2-2g-n)\,\omega_{g,n}(p_1,\dots,p_n)
\eeq
where $\Phi$ is such that $d\Phi=ydx=\omega_{0,1}$.
\et
Notice that this theorem was used to define ${\cal F}_g$ in def \ref{defFg}.

\medskip

$\bullet$ There are many other properties.

For instance the $\omega_{g,n}$'s behave well under taking limits of singular spectral curves, in some sense they commute with taking limit. See \cite{EOFg} for details.

They are also deeply related to integrable systems \cite{BEint, EOFg, EOreview}, and they have many other beautiful properties.

\section{Link to integrability, quantum curves and Hitchin systems}

It is conjectured that the invariants $\omega_{g,n}$ provide the "quantization" of the spectral curve \cite{GSq, MS:2012}.
What we mean by this is deeply related to the notion of integrable systems \cite{BBT,Kri}, and particularly Hitchin's systems and their generalizations.

First let us make some definitions.
We have seen that the tangent space to the space of spectral curves is the space of meromorphic forms. Let $\Omega$ a section of the meromorphic sheaf, and consider the exponential $\e^{\partial_\Omega}$ of the flow $\partial_\Omega$.
We shall often denote:
\beq
{\cal S}+\Omega := \e^{\partial_\Omega}*{\cal S}.
\eeq
By definition, and using the form--cycle duality, we have (this is he Taylor expansion)
$$
\e^{\partial_\Omega}.\omega_{g,n}(p_1,\dots,p_n) = \sum_{k=0}^\infty \frac{1}{k!}\,\oint_{\Omega^*}\dots \oint_{\Omega^*} \omega_{g,n+k}(p_1,\dots,p_k,.,\dots,.)
$$

\bd
We define the formal Tau function:
\beq
{\cal T}_{\hbar}({\cal S}) = \e^{\sum_{g=0}^\infty \hbar^{2g-2}\,{\cal F}_g({\cal S})}
\eeq
and the formal "Baker-Akhiezer" spinor kernel
\beq
\psi({\cal S},\hbar;p,q) =
\frac{\e^{\frac{1}{\hbar}\int_q^p ydx}}{E(p,q)}\,\,\e^{\sum_{2g-2+n>0} \frac{\hbar^{2g-2+n}}{n!} \int_q^p\dots \int_q^p \omega_{g,n}}
\eeq
where $E(p,q)$ is the prime form (defined by $d_{p}d_q \ln{E(p,q)} = B(p,q)$).
For short we shall often write (but keeping in mind that the $(g,n)=(0,2)$ term needs to be regularized)
\beq
\psi({\cal S},\hbar;p,q) \,\,\, "=" \,\,\,
\e^{\sum_{g,n} \frac{\hbar^{2g-2+n}}{n!} \int_q^p\dots \int_q^p \omega_{g,n}}
\eeq
\ed

Almost by definition, and from the form--cycle duality property, it obeys:
\bt[Sato's formula] \cite{EOFg}
\beq
\psi({\cal S},\hbar;p,q) = \frac{{\cal T}_\hbar({\cal S}+\hbar \,\omega_{p,q})}{{\cal T}_{\hbar}({\cal S})}
\eeq
where $\omega_{p,q}^*=[p,q]$, i.e. $\omega_{p,q}(p') = \int_{p''=q}^p B(p',p'')$, i.e. $\omega_{p,q}$ is the 3rd kind differential having a simple pole at $p$ with residue $+1$ and a simple pole at $q$ with residue $-1$.

\et

It is conjectured (under some assumptions which we skip in this short review, see \cite{BEint}) that
\begin{conjecture}
The formal Baker-Akhiezer should be self replicating:
\beq
\delta_{p} \psi({\cal S},\hbar;p_1,p_2) = - \,\psi({\cal S},\hbar;p_1,p)\,\psi({\cal S},\hbar;p,p_2)
\eeq
where the derivation operator $\delta_p$ is the derivative with the flow $\delta_p = dx(p)\,\partial_{\Omega_p}$ along the meromorphic form $\Omega_p(q) = \frac{B(p,q)}{dx(p)}$.

The self-replication formula is in fact equivalent to the Hirota equation for the Tau-function. It also implies many determinantal formulae and Pl\"ucker relations, and the existence of an isomonodromic integrable system, i.e. the existence of an operator $L_q(\hbar, x)\in {\cal M}({\cal C}_0)[[\hbar]]$ (i.e. a formal power series of $\hbar$ whose coefficients are meromorphic 1-forms on the base curve ${\cal C}_0=x({\cal C})$), such that
\beq
(d_p-L_q(\hbar,x(p))) .\psi({\cal S},\hbar;p,q) = 0.
\eeq

\end{conjecture}
This conjecture was never proved in full generality, but has been proved case by case for many families of spectral curves.
The first example was the curve $y=\sqrt{x}$ for which the Baker-Akhiezer kernel is the Airy kernel, proved in \cite{BEdet}.

More examples were proved in \cite{MS:2012, Zhu:2011, BBElax}.

\br
The conjecture as stated above is incomplete. It corresponds only to the case where the curve ${\cal C}$ has genus $\genus=0$.
The full conjecture when the genus $\genus>0$, is stated in \cite{BEint}, it requires to modify the Tau-function and the Baker-Akhiezer kernel with some appropriate Theta--functions (first introduced in \cite{Eyntheta1, EMtheta}), but this is beyond the scope of this short review.

The conjecture was checked for general algebraic spectral curves up to order $O(\hbar^3)$ in \cite{BEint}.
\er

An example where this conjecture is proved in \cite{DimMul13} concerning Hitchin's systems:
\bt[Quantum curves of rank 2 Hitchin systems]
Let ${\cal C}_0$ be a base curve of genus $\genus_0>1$, and ${\cal C}\subset T^*{\cal C}_0\to {\cal C}_0$ be an algebraic curve embedded in the cotangent bundle over ${\cal C}_0$. Let $\eta$ be the restriction of the tautological 1-form on ${\cal C}$.
Consider the spectral curve
$$
{\cal S}=({\cal C},x,y,B)
$$
where $x:{\cal C}\to {\cal C}_0$ is the projection map to the base, $ydx=\eta$ is the tautological 1-form of the cotangent bundle restricted to ${\cal C}$, and $B$ is the canonical 2nd kind differential on ${\cal C}$ (the ${\cal A}$--cycles on which it is normalized are those which projects to the ${\cal A}$--cycles of ${\cal C}_0$ or to $0$, see discussion in \cite{DimMul13}).

Then the Baker-Akhiezer function
\beq
\psi({\cal S},\hbar;p,q) =
\frac{\e^{\frac{1}{\hbar}\int_q^p \eta}}{E(p,q)}\,\,\e^{\sum_{2g-2+n>0} \frac{\hbar^{2g-2+n}}{n!} \int_q^p\dots \int_q^p \omega_{g,n}}
\eeq
turned into a vector by chosing $p_i\in{\cal C}$ the preimages of a point $x\in{\cal C}_0$ of the base by the projection map
\beq
\vec\psi({\cal S},\hbar;x,q) = (\psi({\cal S},\hbar;p_i(x),q))^t \qquad {\rm where}\,x^{-1}(x) = \{p_1(x),\dots,p_r(x)\}
\eeq
 satisfies an isomonodromic ODE
\beq
(\hbar\, d_x-L_q(\hbar,x)) .\vec\psi({\cal S},\hbar;x,q) = 0.
\eeq
where $L_q$ is a Higgs field, whose spectral curve is ${\cal C}$ (the eigenvalues of $L_q$ are independent of $q$, only the eigenvectors depend on $q$).

\et

The generalization to higher rank $>2$ Hitchin systems is at the moment in progress.

\section{A new application: knot theory}

A surprising new application of those invariants and topological recursion to knot theory has emerged recently. Since it attracts lots of activity, and naturally extends the famous (and unproved) Volume conjecture, we mention it here.

\smallskip

Let $\mathfrak{K}$ be a knot embedded in the 3 dimensional sphere $S^3$.
We shall assume here that $\mathfrak{K}$ is a hyperbolic knot, i.e. that $S^3\setminus \mathfrak{K}$ possesses a complete hyperbolic metric (constant curvature $-1$, and such that the total volume is finite). Our favourite example is the figure of eight knot:
$$\includegraphics[scale=0.2]{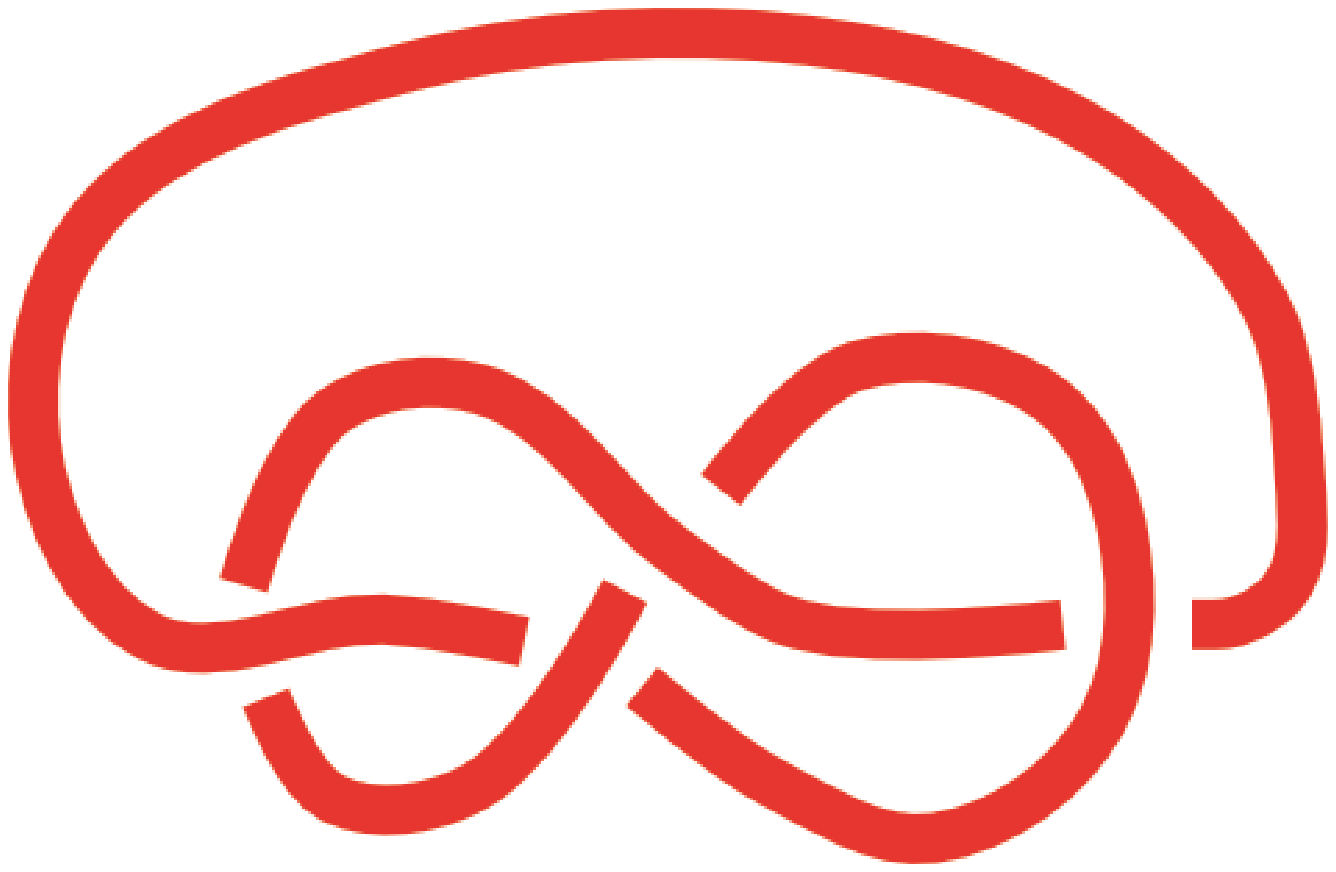}$$

Jones polynomial \cite{Jonesini} is a knot invariant associated to the $SL_2(\mathbb C)$ group.
It is a polynomial of a formal variable $q$, and depends on a representation of $SL_2(\mathbb C)$.
Irreducible representations of $SL_2(\mathbb C)$ are labeled by partitions $\lambda=(\lambda_1,\lambda_2)$ with two rows, and we denote $N=(\lambda_1-1)-(\lambda_2-2)$.
Thus the Jones polynomial $J_N(\mathfrak{K},q)$ is
$$
J_N(\mathfrak{K},q) \in \mathbb Q[q]
\qquad , \quad (N-1,0)={\rm representation\,of}\,SL_2(\mathbb C).
$$

One is often interested in the $q\to 1$ limit together with the large $N$ limit such that
$$
q\to 1\,\, , \quad
N\to\infty\,\, , \quad
u=N\ln{q}={\rm fixed}.
$$
One observes that in that limit the Jones polynomial behaves asymptotically like
\beq
\ln J_n(\mathfrak{K},q) \sim \sum_{k=-1}^\infty \,(\ln q)^k\,\,S_k(u).
\eeq
In 1994, Kashaev \cite{Kash96} made the volume conjecture:
\beq
S_{-1}(2\pi\ii) = {\rm Hyperbolic\, Volume}(S^3\setminus \mathfrak{K}).
\eeq
and this was then generalized  \cite{GuMu06, Murakami1, Hikami, Kashtor, Murarev,Murav} as
\beq
S_{-1}(u) =  {\rm Chern\,Simons\,action}(u).
\eeq
This conjecture was proved for very few knots.

In 2010, Dijkgraaf, Fuji and Manabe \cite{DiFuji1, DiFuji2} conjectured that all $S_k$'s can be found from the topological recursion, in terms of a spectral curve associated to the knot.
The spectral curve is the so-called A-polynomial of the knot, it is also called the character variety of the knot.
For example: the (geometric component) of the A-polynomial of the figure of eight knot has equation:
\beq
\e^{2x}-\e^{x}-2-\e^{_-x}+\e^{-2x} = \e^{y}+\e^{-y}
\eeq
i.e. $x$ and $y$ are (logs of ) meromorphic functions on the algebraic curve ${\cal C}$ of equation $X^2-X-2-X^{-1}-X^{-2}=Y+Y^{-1}$, which is a torus. The fundamental 2nd kind for $B$ is expressed in terms of the Weierstrass function and the second Eisenstein series $E_2$ on this torus:
\beq
B(z,z') = \left( \wp(z-z')\,-\frac{E_2}{3}\right)\,dz\,dz'
\eeq

\begin{conjecture}

The colored  Jones polynomial is a formal Baker-Akhiezer kernel of the spectral curve ${\cal S}$ defined by the A-polynomial:
\beq
J_N(q)^2 = \e^{\sum_{g,n} \frac{(\ln q)^{2g-2+n}}{n!} \int_{D}\dots \int_D \omega_{g,n}}
\eeq
where $D$ is a divisor $\sum_{i=1}^2 p_i -\infty_i$ where $x(p_i)=(\lambda_i-i+c)\ln{q}$ are related to the representation $(\lambda_1,\lambda_2)$, and $c$ is an appropriate constant.
Notice that $u=x(p_1)-x(p_2)=N\ln q$.

The actual statement in fact involves to complete this by the theta terms of \cite{Eyntheta1,EMtheta,BEknot, BEint}.

\end{conjecture}

This conjecture is compatible with the volume conjecture, indeed the leading term:
\beq
S_{-1} = \int_\infty^{p_1} ydx+\int_\infty^{p_2}ydx
\eeq
is the volume.

For the figure of eight knot, this conjecture has been checked up to the 3rd power of $\ln q$, in \cite{BEknot}.

\section{Conclusion}

We hope to have shown the reader that topological recursion is a beautiful and powerful piece of mathematics.
It defines new invariants associated to "spectral curves".

Topological recursion has found a large number of (sometimes unexpected) applications.

However, for many cases, the fact that a given enumerative geometry problem satisfies the topological recursion, is most often only conjectured, not yet proved, and finding proofs is a challenge.
Even in proved cases, the proofs are always very technical and not natural, almost never bijective, so unsatisfactory. Finding a good deep geometric reason (in fact an A-model proof) for the topological recursion is a challenging open problem.

\section*{Aknowledgements}

The author thanks
the CRM Centre de Recherches Math\'ematiques de Montr\'eal,
the FQRNT grant from the Quebec government,
Piotr Su\l kowski and the ERC starting grant Fiels-Knots, and the ICM and Korea.

%\begin{appendices}
%\section{\label{AppendixA} essai \eqref{polestructure}}

%\end{appendices}

\end{document}